\documentclass[twocolumn,superscriptaddress,aps,prx,amssymb,amsfonts,amsmath,floatfix,nofootinbib]{revtex4-1}
\usepackage{bm}
\usepackage{xcolor}
\usepackage{hyperref}
\hypersetup{hypertexnames=false}
\usepackage{graphicx}
\usepackage{float}
\usepackage[normalem]{ulem}
\usepackage[english]{babel}

\let\a=\alpha \let\b=\beta

\let\D=\Delta

\def\aaa{{\bf a}}

\def\to{\rightarrow}  

\newcommand{\beq}{\begin{equation}} \newcommand{\eeq}{\end{equation}}

\def\fth{{\rm th}}

\def\Enoep{{E^{(1)}_\ph}}
\def\ph{{f}}

\begin{document}

\title{Functional bottlenecks can emerge from non-epistatic underlying traits}

\author{Anna Ottavia Schulte}%
\affiliation{Dipartimento di Fisica, Sapienza Universit\`a di Roma, Piazzale Aldo Moro 5, 00185 Rome, Italy}

\author{Samar Alqatari}
\affiliation{Department of Physics and The James Franck and Enrico Fermi Institutes,
The University of Chicago, Chicago, IL 60637, USA}

\author{Saverio Rossi}%
\affiliation{Dipartimento di Fisica, Sapienza Universit\`a di Roma, Piazzale Aldo Moro 5, 00185 Rome, Italy}

\author{Francesco Zamponi}%
\affiliation{Dipartimento di Fisica, Sapienza Universit\`a di Roma, Piazzale Aldo Moro 5, 00185 Rome, Italy}

\begin{abstract}
Protein fitness landscapes frequently exhibit epistasis, where the effect of a mutation depends on the genetic context in which it occurs, \textit{i.e.}, the rest of the protein sequence. Epistasis increases landscape complexity, often resulting in multiple fitness peaks.
In its simplest form, known as global epistasis, fitness is modeled as a non-linear function of an underlying additive trait. In contrast, more complex epistasis arises from a network of (pairwise or many-body) interactions between residues, which cannot be removed by a single non-linear transformation.
Recent studies have explored how global and network epistasis contribute to the emergence of functional bottlenecks - fitness landscape topologies where two broad high-fitness basins, representing distinct phenotypes, are separated by a bottleneck that can only be crossed via one or a few mutational paths.
Here, we introduce and analyze a stylized model of global epistasis with an additive underlying trait. We demonstrate that functional bottlenecks arise with high probability if the model is properly calibrated. Furthermore, our results underscore that a proper balance between neutral and non-neutral mutations is needed for the emergence of functional bottlenecks. 
\end{abstract}

\maketitle

\subsection*{Author Summary}
A central challenge in the study of protein evolution is understanding how interactions between mutations influence evolutionary dynamics. These interactions, collectively known as epistasis, play a key role in shaping protein 'fitness landscapes', representing maps between amino acid sequences and their functional performance. However, the impact of epistasis on landscape ruggedness and accessibility remains a subject of debate.

Recent experiments have revealed the existence of functional bottlenecks: regions of low fitness that restrict evolutionary transitions between proteins with different functionalities. While these bottlenecks represent a significant constraint on evolutionary outcomes and are often attributed to complex networks of interacting mutations, we demonstrate that they can arise in a much simpler setting.

Using a stylized model where fitness depends nonlinearly on an underlying additive trait, we show that functional bottlenecks do not require complex network epistasis. Instead, they can emerge from the inherent variability of the mutational effect distribution. Specifically, these bottlenecks arise when the mutational landscape is dominated by small, nearly neutral effects, but remains punctuated by enough strongly non-neutral mutations to create sharp fitness transitions. Our findings offer a new perspective on the constraints shaping protein evolution and highlight the critical role of mutational effect heterogeneity in determining evolutionary accessibility.

\section{Introduction}

Understanding how genetic variation translates into differences in reproductive success is a central challenge in evolutionary biology. An organism’s fitness — defined as the number of offspring it produces — depends on how its phenotypic traits interact with the environment. These traits, in turn, are shaped by the organism’s genotype, suggesting a hierarchical mapping from genotype to phenotype to fitness. 
By assigning a fitness value to each genotype, one obtains a {\it fitness landscape}~\cite{wright1932roles,de2014empirical,Tenaillon2014,Manrubia2021,Pahujani2025}, a conceptual tool that captures the topology of the space of evolutionary possibilities. 

Epistasis is a distinctive property of fitness landscapes.
Broadly speaking, it is defined as the context-dependence of mutational effects~\cite{Bateson1907,Fisher1919,lunzer2010pervasive,harms2013evolutionary,weinreich2013should,olson2014comprehensive,figliuzzi2016coevolutionary,starr2016epistasis,poelwijk2016context,rivoire2016evolution,miton2016mutational,sailer2017high,sailer2017detecting,Hwang2017,cocco2018inverse,domingo2018pairwise,starr2018pervasive,otwinowski2018biophysical,otwinowski2018inferring,domingo2019causes,Biswas2019,poelwijk2019learning,ballal2020sparse,lyons2020idiosyncratic,miton2021epistasis,reddy2021global,phillips2021binding,bakerlee2022idiosyncratic,vigue2022deciphering,chen2024understanding,johnson2023epistasis,buda2023pervasive,alqatari2024epistatic}. 
More precisely, it is defined as follows. 
Consider a genotype $\aaa = (a_1,a_2,\cdots,a_L)$ with an associated fitness $F(\aaa)$.
The genotype $\aaa$ could be a protein sequence of length $L$, with $a_i$ representing the amino acid at site $i$, or a nucleotide sequence, or a sequence of zeros and ones representing the presence/absence of a given mutation, gene, etc.
Epistasis occurs when the fitness change $\Delta F_i(a\to b)$ due to substituting $a_i=a$ with $a_i=b$ at site $i$ depends on the amino acids %$(a_1,\cdots,a_{i-1},a_{i+1},\cdots, a_L)$ 
that are present at other sites in the sequence.
As such, epistasis can be of two conceptually distinct origins. 

What is sometimes called `global' epistasis is based on introducing the simplest possible non-linearity in the genotype-phenotype-fitness mapping~\cite{de2014empirical,Tenaillon2014,Pahujani2025,Hwang2017,otwinowski2018biophysical,otwinowski2018inferring,poelwijk2019learning,husain2020physical,reddy2021global,phillips2021binding,johnson2023epistasis,buda2023pervasive}.
More precisely, one assumes the existence of an underlying additive phenotype (or `trait') $E(\aaa) = \sum_{i=1}^L h_i(a_i)$ associated with each genotype $\aaa$. 
Under this assumption, the variation of phenotype $E(\aaa)$ associated with substituting $a$ with $b$ at site~$i$, $\Delta E_i(a\to b) = h_i(b) - h_i(a)$, is independent of the rest of the genotype.
Next, one assumes that the fitness $F(\aaa) = \phi(E(\aaa))$ is a non-linear function of $E(\aaa)$, for instance a sigmoid (Fig.~\ref{fig:1}a).
Under these assumptions, while the variation of the underlying trait is independent of the rest of the sequence, the variation of fitness depends on it due to the non-linearity of $\phi$.
Yet, such non-linearity is relatively easy to handle, because one can deduce the function $\phi(E)$ by `deconvolution' of experimental fitness measurements~\cite{otwinowski2018biophysical,otwinowski2018inferring,poelwijk2019learning,phillips2021binding,buda2023pervasive,park2024simplicity,dupic2024protein,carlson2025distinguishing}, and thus describe the fitness function with a limited number of parameters. 
In concrete examples, few parameters enter in the definition of $\phi(E)$, and of the order of $L$ parameters enter in the definition of $E$. 
Furthermore, such a globally epistatic fitness landscape only features a single maximum, because $F=\phi(E)$ is an increasing function of $E$, and $E$ itself has a single maximum, obtained by choosing at each site the value $a_i$ that maximizes the additive contribution $h_i(a_i)$~\cite{Pahujani2025}. Multi-peaked landscapes can be obtained
by choosing a non-monotonic function $\phi(E)$, as in Fisher's Geometric Model~\cite{de2014empirical,Tenaillon2014,martin2015fitness,Hwang2017,Pahujani2025}.

A more complex behavior is obtained in presence of what is sometimes called `network' epistasis~\cite{lunzer2010pervasive,olson2014comprehensive,figliuzzi2016coevolutionary,rivoire2016evolution,poelwijk2016context,cocco2018inverse,starr2018pervasive,ballal2020sparse,russ2020evolution,lyons2020idiosyncratic,miton2021epistasis,bakerlee2022idiosyncratic,vigue2022deciphering,chen2024understanding,park2024simplicity,dupic2024protein,carlson2025distinguishing}, which
implies that no function $E=\phi^{-1}(F)$ can transform fitness into a purely additive function of the genotype~\cite{poelwijk2019learning,phillips2021binding,buda2023pervasive}.
While a function $\phi$ can be chosen to make $E$ as close as possible to an additive form, higher-order interactions will still play a role in the presence of network epistasis.
More formally, $E$ can always be decomposed as a sum of additive, two-site, three-site, and higher-order interactions~\cite{weinreich2013should,poelwijk2016context,sailer2017detecting,sailer2017high,domingo2018pairwise,poelwijk2019learning,ballal2020sparse,phillips2021binding,buda2023pervasive}, i.e.,
\beq\label{eq:Edec}\begin{split}
E(\aaa) =& \sum_{i} h_i(a_i) + \sum_{i,j} J_{ij}(a_i,a_j) \\& + 
\sum_{i,j,k} K_{ijk}(a_i,a_j,a_k) + \cdots
\end{split}\eeq
however, due to network epistasis, no choice of $\phi$ can eliminate the higher-order terms $J_{ij}$, $K_{ijk}$, and beyond. 
Let us focus for simplicity on the case in which only the two-site interactions are present. 
The number of parameters required to fully specify the function $E$ now scales as the number of pairs, i.e., proportionally to $L^2$, which makes inferring the fitness function from a few measurements significantly more difficult.
Even more importantly, the function in Eq.~\eqref{eq:Edec} can have many local maxima, corresponding to many fitness peaks, leading to a rough, complex fitness landscape.
From Eq.~\eqref{eq:Edec}, any nonlinear function of $E(\aaa)$ -- that is, any form of global epistasis -- can, in principle, be represented as a (possibly infinite) sum of high-order terms on the right-hand side.
However, this representation is less practical, as it requires estimating a large number of parameters. 
A more effective strategy is to first capture as much of the nonlinearity as possible using a single global function. 
This substantially reduces the residual nonlinearity, making it easier to model the remainder with only low-order local terms. 

The relative importance of global and network epistasis, and how to properly fit $E(\aaa)$, has recently been the subject of intense debate~\cite{park2024simplicity,dupic2024protein,carlson2025distinguishing}.
A particular focus of the debate has been the role played by epistasis in shaping the topology of the fitness landscape in the proximity of functional switches.
For instance, Poelwijk et al.~\cite{poelwijk2019learning} studied experimentally the fitness landscape separating two phenotypically distinct variants of the {\it Entacmaea quadricolor} fluorescent protein, one fluorescent in the red and the other in the blue, separated by 13 mutations.
Assaying all $2^{13}$ intermediate variants, they observed, in addition to global epistasis, significant network contributions (both pairwise and higher order) to fitness.
They also identified a functional bottleneck separating the `blue' and `red' space of fluorescent variants, i.e. a narrow region of accessible evolutionary paths between the two phenotypes, which they attributed to epistatic constraints.
These functional bottlenecks illustrate how fitness landscapes can remain navigable despite their ruggedness. While epistasis can constrain viable evolutionary paths, it can also facilitate adaptation by carving out narrow high-fitness pathways that connect otherwise isolated functional states -- typically via critical “switch” mutations~\cite{weinreich2006darwinian, wu2016adaptation, kantroo2024high}.

In a subsequent study, Alqatari and Nagel~\cite{alqatari2024epistatic} used similar methods to investigate functional switches in an elastic network, considered as a model for allosteric behavior in proteins. Analyzing ensembles of fitness landscapes, they observed different topologies at the threshold for viable evolution, and found that functional bottlenecks -- often involving a single critical mutation -- emerged generically, similar to the observations in the experimental study.

Many other recent studies investigated small combinatorial fitness landscapes. However, in most cases, these landscapes connect a reference wild type to an evolved variant with increased or decreased fitness with respect to a single functionality~\cite{buda2023pervasive, papkou2023rugged, johnston2024combinatorially}. Experimental studies that investigate the combinatorial landscape separating two reference genotypes with distinct functionality are less widespread~\cite{westmann2024entangled, starr2017alternative, Hayden2019ribozymes, phillips2021binding}. Moreover, in many of these studies functional transitions proceed through promiscuous intermediate variants that are as fit, or even fitter, than the two reference genotypes. 
Allowing for functional promiscuity represents a less constrained scenario than one where a functional switch must occur between mutually exclusive phenotypes. In the latter case, which is the one we consider here, the requirement for specialization naturally makes the landscape more prone to the formation of bottlenecks.
The question of how widespread bottlenecks are -- particularly in presence of promiscuity -- is not addressed here.

In this work, we aim to identify the minimal conditions under which bottleneck topologies can emerge. Specifically, we investigate whether network epistasis is a {\it necessary} condition for the existence of functional bottlenecks in fitness landscapes. To address this, we introduce a simple, stylized model of global epistasis based on a random underlying additive trait and a nonlinear fitness function. Our model is not intended to fit specific empirical data; rather, it follows the spirit of the Fisher's Geometric Model and related frameworks~\cite{de2014empirical,Tenaillon2014,martin2015fitness,Hwang2017,Pahujani2025} by generating an ensemble of random fitness landscapes with stylized features. Upon calibration, we demonstrate that our model encompasses a wide range of parameters for which the resulting topologies exhibit a functional bottleneck with high probability.
This result, by itself, shows that global epistasis alone is sufficient to generate functional bottlenecks.

Furthermore, we studied the role of the distribution of single mutational effects (SMEs). 
More specifically, we find that when the model is properly calibrated, the distribution of SMEs selected by evolution displays a proper balance between nearly neutral SMEs and enough strongly non-neutral SMEs, which is needed to
create sharp fitness transitions.
Our stylized model thus sheds light on some fundamental constraints that shape fitness landscapes and possibly evolution. 

\section{Results}

The central result of our work is the introduction of a stylized fitness landscape model, designed to investigate the minimal conditions under which bottleneck topologies emerge. However, to provide the motivation and inspiration for this construction, we begin with a pedagogical discussion of experimental data from Poelwijk et al.~\cite{poelwijk2019learning}. This analysis of empirical data serves to ground the stylized features of the model that follows.

\subsection{Experimental data on fitness bottlenecks}
\label{sec:exp}

\begin{figure*}[t]
\centering
    \includegraphics[width=\linewidth]{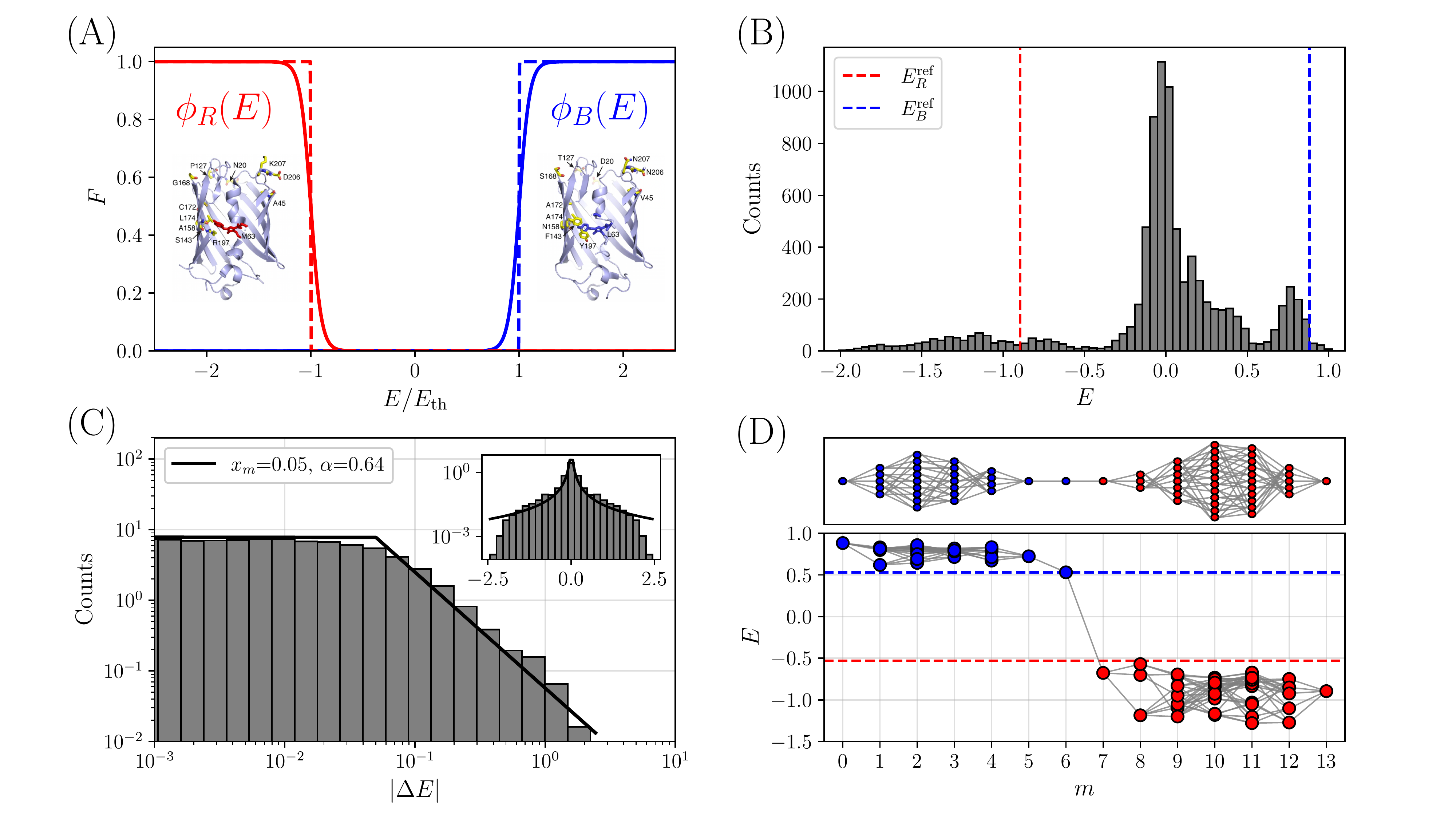}
\caption{
Experimental data from Ref.~\cite{poelwijk2019learning}.
(A)~Schematic shape of the fitness functions for the red and blue phenotypes as a function of the underlying trait $E$. 
Sigmoid functions have been used for illustrations. 
The two reference structures with the corresponding mutations are also shown, from Ref.~\cite{poelwijk2019learning}.
(B)~Histogram of the value of $E$ obtained from Eqs.~(\ref{eq:EvsF},\ref{eq:EAEB}) for each of the $2^{13}$ variants. 
The red and blue lines correspond respectively to the reference values $E^{\rm {ref}}_R$ and $E^{\rm {ref}}_B$. 
(C)~Histogram of the absolute value $|\D E|$ of single mutational effects (SMEs) from Eq.~\eqref{eq:SME} for all $13\times 2^{13}$ single mutations that can be obtained from the dataset, shown in log-log (main panel) and lin-log (inset) scales. 
The black curve is the fit obtained with the Pareto distribution in Eq.~\eqref{eq:Pareto}.
(D)~Topology of the space of paths obtained keeping only genotypes $\aaa$ with $|E(\aaa)|>E_C$, and finding the largest possible value of $E_C=0.53$ (dashed line) such that the red and blue reference sequences remain connected. 
The lower panel reports the values of $E(\aaa)$ for each functional genotype (blue dots for $E>E_C$ and red dots for $E<-E_C$) as a function of the number of mutations $m$ from the blue reference, with the gray lines connecting pairs of genotypes that differ by a single mutation. 
The upper panel shows the resulting graph of connections.
}
\label{fig:1}
\end{figure*}

Poelwijk et al.~\cite{poelwijk2019learning}
analyzed a family of variants of
the naturally occurring wild type \textit{Entacmea Quadricolor} red fluorescent protein \textit{eqFP611}~\cite{wiedenmann2005red}.
More specifically, from a natural variant of \textit{eqFP611}, called \textit{eqFP578} (with 76\% amino acid sequence homology),
the \textit{TagRFP} mutant was engineered via random mutagenesis.
\textit{TagRFP} is 21 mutations away from \textit{eqFP578}~\cite[see SI]{merzlyak2007bright} and has enhanced fluorescence and stability.
From \textit{TagRFP}, two
engineered variants have been derived to have distinct fluorescence\footnote{See {\tt https://www.fpbase.org/protein/eqfp578} for the complete tree of variants.}: 
\begin{itemize}
\item \textit{mTagBFP2}, which is fluorescent in the blue and has 13 mutations and 6 insertions relative to \textit{TagRFP}~\cite{subach2011enhanced};
\item 
\textit{mKate2}, which is fluorescent in the deep red and has 8 mutations and one insertion relative to \textit{TagRFP}~\cite{shcherbo2009far}.
\end{itemize}
The sequences of \textit{mTagBFP2} and of \textit{mKate2}
differ by 13 mutations (12 of which are shown in Fig.~\ref{fig:1}a reproduced from Ref.~\cite{poelwijk2019learning}), thus resulting in a total of $2^{13} = 8192$ different combinations of intermediate amino acid substitutions.
The experiment of Poelwijk et al.~\cite{poelwijk2019learning} measured the fluorescence of all these intermediates, both in the red and in the blue.
Note that there are also 5 insertions in \textit{mTagBFP2} with respect to \textit{mKate2} that were neglected in the experiment (details were not found in Ref.~\cite{poelwijk2019learning}).

We focus on these published data as they characterize a combinatorial landscape bridging two reference genotypes with distinct functionalities, within which a functional bottleneck was explicitly identified. A more recent study, employing a similar experimental design but investigating a different protein, reported the absence of such a bottleneck~\cite{westmann2024entangled}. This discrepancy highlights the topological variability of fitness landscapes and motivates our search for the minimal conditions that govern the emergence of these constraints.

We downloaded the original data from Ref.~\cite{poelwijk2019learning} and represented the genotypes with binary vectors of 13 variables, with $\aaa_B=(0,0,\cdots,0)$ being the blue \textit{mKate2} and $\aaa_R=(1,1,\cdots,1)$ being the red \textit{mTagBFP2}. In the experiment, {\it E.Coli} cells expressing the mutant genotypes were sorted by a microfluidic device able to push them into separate channels according to their fluorescence.
From the raw experimental measurement, we assign to each genotype a red fitness, $F_R(\aaa)$, and a blue fitness, $F_B(\aaa)$, based on the enrichment in the two channels, each normalized to its corresponding reference value such that $F_R(\aaa_R)= F_B(\aaa_B)=1$ (see the SI for details). This normalization provides a consistent reference scale, with each reference variant serving as the baseline functional state for its phenotype, enabling a more direct quantitative comparison between the red and blue fitness values. 

Note that a specific measurable trait -- fluorescence intensity -- is here employed as a proxy for fitness, despite its lack of direct coupling to reproductive success. Instead, fluorescence quantifies the protein’s functional efficiency. Under the assumption that enhanced biophysical performance may confer a selective advantage within the relevant environment, fluorescence serves as a proxy for fitness. This approach is standard in many empirical studies of fitness landscapes, such as Refs.~\cite{poelwijk2019learning,westmann2024entangled}.

For each measured phenotype (we will use a suffix `R' for red and `B' for blue fluorescence), we can define an underlying trait and a non-linear mapping to fitness ${F=\phi(E)}$. 
Inverting the non-linear function, we can thus derive the underlying trait from the measured fitness,
\beq\label{eq:EvsF}
E_R(\aaa) = \phi_R^{-1}\left[F_R(\aaa)\right] \ , \quad
E_B(\aaa) = \phi_B^{-1}\left[F_B(\aaa)\right]  \ .
\eeq
The optimal methodology for inferring the function $\phi(E)$ remains a subject of active debate~\cite{otwinowski2018inferring,phillips2021binding,buda2023pervasive,park2024simplicity,dupic2024protein,carlson2025distinguishing}; for instance, it has been suggested that the nonlinearity be inferred directly through rank-based statistics~\cite{carlson2025distinguishing}. 
While acknowledging that results can be sensitive to the specific choice of $\phi$, for our illustrative purposes we keep the same power-law form $\phi^{-1}(x)=x^{0.44}$ utilized in Ref.~\cite{poelwijk2019learning} for both phenotypes. Alternative specifications, such as the sigmoid function shown in Fig.~\ref{fig:1}a, may yield quantitatively different results, but the qualitative topology of the landscape remains robust.

We note that when there are two phenotypes only, for simplicity, we can follow Alqatari and Nagel~\cite{alqatari2024epistatic} and
encode them both in a single trait
\begin{equation}\label{eq:EAEB}
E(\aaa) = E_B(\aaa) - E_R(\aaa) \ ,
\end{equation}
which is positive for the `blue' phenotype and negative for the `red' phenotype.
This is only possible when the two phenotypes are exclusive, as in Ref.~\cite{poelwijk2019learning}, which may not be true in other cases~\cite{Hayden2019ribozymes,phillips2021binding,westmann2024entangled,starr2017alternative}.
We also define the reference trait values as
\beq
E^{\rm {ref}}_B = E(\aaa_B) \ , \qquad E^{\rm {ref}}_R = E(\aaa_R) \ .
\eeq
The structure of the two proteins and the 13 substitutions are illustrated in Fig.~\ref{fig:1}a, together with a schematic illustration of the non-linear mapping between $E=E_B-E_R$ and the two fitnesses. 

Fig.~\ref{fig:1}b shows the histogram of $E(\aaa)$ across all $2^{13}$ genotypes, with red and blue lines corresponding to the two reference variants. 
Note that the reference values are $E^{\rm {ref}}_B \sim 0.88$ and $E^{\rm {ref}}_R \sim -0.89$.
Their absolute value slightly differs from one due to the contribution from the other phenotype.
As observed in Ref.~\cite{poelwijk2019learning}, most genotypes are non-functional with $E\sim 0$.
There is a rather sharp peak of blue genotypes around $E \sim 1$, while red genotypes exhibit a much broader distribution, ranging from $E \sim -0.5$ down to large negative values of $E$.
This indicates that some intermediates are significantly more fluorescent in the red than the red reference variant itself.

Having defined $E(\aaa)$, we can analyze the distribution of SMEs. 
For each of the $2^{13}$ backgrounds $\aaa$, we consider all 13 single mutations and we compute the SME,
\beq\label{eq:SME}
\Delta E_i(\aaa) = E(a_i \to 1-a_i \, | \aaa) - E(\aaa) \ .
\eeq
The histogram of the absolute value of the $13\times 2^{13}$ SMEs is reported in Fig.~\ref{fig:1}c. (Note that the $13\times 2^{13}$ SMEs include each mutation and its reverse, hence the histogram would be symmetric by construction, which is why we focus here on $|\Delta E_i|$.)
Although we have limited data, we clearly observe that the distribution is fat-tailed. 
A good fit is achieved by a Pareto distribution, 
\begin{equation}\label{eq:Pareto}
    p(x) = \frac{\a}{2x_m(\a+1)} \times
    \begin{cases}
    1& \text{if } |x|<x_m  \ , \\
    \big(\frac{x_m}{|x|}\big)^{\alpha+1}& \text{if } |x|>x_m \ , \\
    \end{cases}
\end{equation}
here with $x=\Delta E$.
The fit is just an indication, in particular due to the limited amount of data and the limited range of the experimental fitness, which both cutoff the distribution at large $\Delta E$.
While other distributions could also fit the data, the distribution tails are quite broad, indicating substantial heterogeneity among SMEs (in the SI we show similar results for another, independent dataset of SMEs~\cite{buda2023pervasive}).

Finally, in Fig.~\ref{fig:1}d, we examine the topology of the resulting functional bottleneck. To perform this analysis, we need to first establish a formal definition of a `functional' genotype. As noted in Ref.~\cite{poelwijk2019learning} and illustrated in Fig.~\ref{fig:1}b, the threshold for functionality is inherently context-dependent; shifting this threshold can substantially alter the perceived landscape topology~\cite{alqatari2024epistatic}. Indeed, viability is not an intrinsic property of a genotype alone but is contingent upon the environment -- specifically the selection pressure exerted on the relevant trait -- which varies across different experimental conditions.
In this work, we consider that the fitness function is growing fast around a given threshold $E_\fth$ (either on the positive or negative side, see Fig.~\ref{fig:1}a). 
Hence, genotypes with $|E(\aaa)| < E_\fth$ are considered dysfunctional and evolutionarily disallowed.
Because the reference variants are functional, they must of course be located above the functionality threshold, i.e. $|E^{\rm {ref}}| \gtrsim E_\fth$. 
However, Fig.~\ref{fig:1}d shows that no path of single mutations exists that connects the two reference variants while always maintaining $|E(\aaa)|\geq E^{\rm {ref}}$, hence we need to consider $E_\fth < E^{\rm {ref}}$ in order to preserve the connection between the reference sequences. (Note that a different result was obtained in Ref.~\cite{poelwijk2019learning} when the two phenotypes were not normalized to the reference values. We discuss this point in the SI.)
What is, then, the largest value of the functionality threshold $E_\fth$ such that at least one viable mutational path exists between the two reference variants? 

In order to precisely answer this question,
following Ref.~\cite{alqatari2024epistatic}, 
we characterize the space of the paths connecting the two reference variants upon increasing the threshold~$E_\fth$.
We consider the subset of paths that (i) are made of single mutations, (ii) connect the red and blue reference sequences, and (iii) are such that $|E(\aaa)| > E_\fth$ all along the path. 
We start from $E_\fth=0$, and gradually increase $E_\fth$ to the maximum value such that at least one path remains. 
The resulting value, which we call $E_C$, quantifies how `hard' it is for evolution to find a path that realizes the functionality switch. 
If $E_C\sim E^{\rm {ref}}$, then one can find a path of single mutations, such that all intermediates are equally functional to the reference variants. 
If instead $E_C \ll E^{\rm {ref}}$, in order to connect the two reference variants one has to accept a loss of fitness with respect to the reference fitness, which makes the transition less likely, but still possible.

For the experimental data of Ref.~\cite{poelwijk2019learning}, this analysis yields a value of approximately $E_C \sim 0.53$, or $E_C/E^{\rm {ref}} \sim 0.6$.
The resulting space of paths, depicted in Fig.~\ref{fig:1}d, exhibits the characteristic bottleneck shape also reported in Ref.~\cite{alqatari2024epistatic}, where all paths traverse a single `jumper' genotype with $E=E_C$. 
In Fig.~\ref{fig:1}d, the jumper genotype has a blue phenotype, after which a single mutation brings to the red phenotype.
This functional switch is driven by a mutation located roughly at the midpoint.
The number of functional intermediates determines the number of evolutionary paths that can reach the critical `jumper' genotype. More paths imply greater evolutionary accessibility, making it easier for evolution to find the key mutation; while fewer paths indicate a more constrained transition requiring a specific mutational sequence.
Note that the analysis in Ref.~\cite{poelwijk2019learning} found a less pronounced bottleneck, with multiple mutational paths surviving -- a scenario that can be reproduced here by adopting a less stringent definition of $E_C$ which allows for multiple paths, see Ref.~\cite{alqatari2024epistatic}.

\subsection{Stylized model of global epistasis}

The goal of this paper is to determine whether a bottleneck structure like the one observed in Sec.~\ref{sec:exp}  requires network epistasis, or if it can also be explained by a simpler model of global epistasis.

To formulate a simple stylized model, we begin by an important observation.
The reference proteins analyzed in Ref.~\cite{poelwijk2019learning} are not the product of natural evolution, but rather highly engineered sequences.
In engineered proteins, derived either by random mutagenesis as in Sec.~\ref{sec:exp} or by directed evolution, one typically observes that the fitness improvement is achieved by just a few highly beneficial 
mutations, see e.g.~\cite{zhong2020automated,buda2023pervasive,frohlich2024epistasis}. 
While this might seem, in principle, a highly specific feature of engineered proteins, a similar scenario has been observed when a natural protein needs to quickly adapt to a novel environment and acquire a new function. A notable example is the SARS-COV-2 Spike protein that displayed a few highly beneficial mutations just after the virus jumped to human hosts (see e.g.~\cite{rodriguez2022epistatic}), but similar dynamics has been observed in other viral proteins~\cite{hadfield2018nextstrain}, and in the immune system~\cite{phillips2021binding}.
In each of these cases, just a few advantageous mutations allow the protein to acquire the new desired function, whether the selection pressure is applied artificially or emerges from a change in the natural environment. A very different dynamics might be at play in the neutral evolution of optimized initial proteins~\cite{fantini2020protein,stiffler2020protein,bisardi2022modeling,DiBari2024}, which might explain why the combinatorial landscape of Ref.~\cite{westmann2024entangled} does not display a bottleneck.

To capture a few minimal ingredients inspired by the results of Sec.~\ref{sec:exp} and the above discussion, we introduce a stylized model featuring only global epistasis. 
The model is defined as follows:
\begin{itemize}
\item The genotype $\aaa = (a_1,a_2,\cdots,a_L)$ is a binary sequence (i.e., $a_i \in \{0,1\}$) of length $L$.
\item An underlying additive trait $E(\aaa) = \sum_{i=1}^L h_i a_i$ has
random SMEs $h_i$ that are identically and independently distributed according to a symmetric input distribution $P(h) = P(-h)$. 
After having drawn the values $h_i$ independently, we impose that $\sum_{i=1}^L h_i=0$ by shifting their mean, i.e., $h_i \to h_i - \frac1L\sum_{i=1}^L h_i$.
\item The two fitness functions that will be associated with the two `colors' are non-linear functions of the underlying trait, $F_B=\phi_B(E) = \phi_0/(1+e^{\b(E_\fth-E)})$ and $F_R= \phi_R(E) = \phi_0/(1+e^{\b(E_\fth+E)})$, as illustrated in Fig.~\ref{fig:1}a.
If $E(\aaa) > E_\fth$ then the genotype $\aaa$ is functional `blue', if $E(\aaa) < - E_\fth$ it is functional `red', and if $|E(\aaa)|\ll E_\fth$, the genotype is non-functional. 
\end{itemize}
The fitness functions grow sharply around $\pm E_\fth$, in a way controlled by the parameter $\beta$.
The choice we will make implicitly, following Ref.~\cite{alqatari2024epistatic}, is to consider the limit $\beta \gg 1$, in which the fitness is close to a threshold (or Heaviside) function at $\pm E_\fth$ (see Fig.~\ref{fig:1}a, dashed curves). 
Note that the values $h_i$ can be multiplied by an arbitrary constant that can be absorbed into $E_\fth$ and $\beta$, which allows us to fix 
the overall scale (e.g. the variance) of $P(h)$ without loss of generality.

Given the above definitions, both genotypes $\aaa = \mathbf{0} = (0,0,\cdots,0)$ and $\aaa = \mathbf{1} = (1,1,\cdots,1)$ have $E=0$ and are thus non-functional. 
We stress that, contrarily to the analysis of experimental data reported above, these two genotypes will not be the reference sequences in the stylized model.
Following Ref.~\cite{alqatari2024epistatic}, we thus introduce a `tuning procedure' to generate two reference variants, the `red' $\aaa_R$ with $E_R^{\rm {ref}} <  -E_T$ and the `blue' $\aaa_B$ with $E_B^{\rm {ref}} >  E_T$, for a chosen `tuning' value $E_T>0$. 

The procedure starts from the genotype $\aaa = \mathbf{0} = (0,0,\cdots,0)$, which is thus considered as a `common ancestor' or `wild type' (for example, in the experiment described in Sec.~\ref{sec:exp}, it would correspond to the \textit{eqFP578} protein).
One then sequentially introduces mutations as follows.
\begin{itemize}
\item With probability $p$, we perform a `greedy' step. 
We scan all the $h_i$ values that have not been introduced yet (those for which $a_i=0$), and choose the one that has the maximum impact towards the desired goal, i.e., the largest positive $h_i$ for the blue reference variant or the smallest negative $h_i$ for the red reference variant. 
We introduce the corresponding mutation by switching $a_i=1$.
\item With probability $1-p$, we perform a `random' step. 
We choose at random one site $a_i$ such that $a_i=0$, and mutate it to $a_i=1$.
\end{itemize}
The tuning procedure is performed independently for the two colors starting from $\aaa=\mathbf{0}$, and continues until for the first time $E^{\rm {ref}}_R <  -E_T$ to construct the red reference variant, or $E^{\rm {ref}}_B >  E_T$ to construct the blue reference variant.

This procedure is not designed to be a fully realistic model of molecular evolution. 
Nonetheless, it is grounded in experimental observations. 
As noted at the beginning of this section, directed evolution experiments often yield final protein variants carrying a small number of beneficial mutations together with a few neutral ones. 
More generally, when a protein is challenged to acquire a new function, adaptation often involves selecting a handful of advantageous substitutions on top of incidental neutral changes. 
Our approach for constructing the reference variants is simply the most straightforward way to reproduce this pattern, without any claim of population-genetic realism.
We also believe that this procedure mimics the tuning of elastic networks performed in Ref.~\cite{alqatari2024epistatic}, in which the network is tuned based on a local proxy for the global fitness; we hereby assume that this local proxy can indeed provide the best choice (with probability $p$) or miss it (with probability $1-p$).

The parameters that define the model are 
the sequence length $L$, the {\it a priori} distribution of SMEs $P(h)$, the probability $p$ and the tuning parameter $E_T$ that controls the reference sequence construction.
We note that $P(h)$ may be interpreted as describing the a priori distribution of mutational effects -- arising, for instance, from underlying biophysical constraints -- which represents the potential 'pool' of mutations accessible to evolution. We found that various choices of $P(h)$ lead to similar outcomes, provided the model is appropriately calibrated. For the sake of parsimony, and recognizing that these are stylized representations, we restrict our analysis to two distributions that capture distinct mutational behaviors.
\begin{enumerate}
\item {\bf Gaussian}: A Gaussian distribution with unit variance.
\item {\bf Pareto cutoff}:
Motivated by the analysis of experimental data in Fig.~\ref{fig:1}c, a Pareto distribution, i.e., Eq.~\eqref{eq:Pareto} with $x=h$, where we fix $\a=0.7$ and $x_m=0.1$.
We also include an upper cutoff as in the data, i.e. we set $P(h)=0$ for $|h|>2$ and adjust the normalization constant accordingly. 
\end{enumerate}

For a given choice of parameters, an `instance' of our random model is thus defined by the {\it a priori} pool of the $L$ SMEs, $h_i$, and by the two reference variants $\aaa_R$ and $\aaa_B$. 
For a given instance, we can define the number of mutations $M$ separating $\aaa_R$ and $\aaa_B$ (i.e., the number of $a_i$ that differ in the two sequences), and we can then investigate the $M!$ directed paths that connect them, obtained by introducing these mutations in all possible orders.
Following Ref.~\cite{alqatari2024epistatic} and as discussed above, we can find the maximum value of $E_\fth$ such that at least one single-mutational path connecting the two reference variants, with all intermediate states having $|E|>E_\fth$,
survives, and we call this value $E_C$. 
The closer $E_C$ is to $E^{\rm {ref}}$, the easier it is to perform the functional switch. 

Given $(L,p,E_T)$ and $P(h)$, we wish to characterize the statistical properties of the quantities $M$, $E^{\rm {ref}}$, and $E_C$, which characterize the overall topology of the fitness landscape. 
Another interesting quantity is the distribution $P(\tilde h)$ of SMEs $\tilde{h}_i$ selected by the tuning procedure to generate the two reference variants (i.e., those $h_i$ that correspond to $a_i=1$ in each reference genotype), which may be interpreted as the distribution of mutations that are `fixed' by evolution.
We will also define $E^{\rm {ref}}_{\rm max} = \max(E^{\rm {ref}}_B,|E^{\rm {ref}}_R|)$.
In the following, we will use brackets, $\langle A\rangle$, to denote the statistical average of an observable $A$ over an ensemble of random instances generated by the model.
Unless otherwise specified, 
the average is taken over 20000 independent instances of the model for each value of the parameters. 

Our model is similar in spirit to Fisher's geometric model (FGM) as discussed in Refs.~\cite{de2014empirical,Tenaillon2014,martin2015fitness,Hwang2017,Pahujani2025}, from which it however differs in two important aspects. 
First, the choice of non-linearity, which reflects here the existence of two distinct fitness functions associated to each color, see also Ref.~\cite{martin2015fitness} for a similar choice. 
Second, the fact that we choose two reference variants using the stochastic procedure described above, and we then restrict to the space of intermediates between the selected variants. 
As we show in the following, these two ingredients, together with a proper calibration of the model, give rise to the desired bottleneck structure.

Before discussing the details of the calibration procedure, a typical example of the space of mutational paths obtained from a calibrated model is shown in Fig.~\ref{fig:2}. 
The qualitative similarity with Fig.~\ref{fig:1}d, modulo the (irrelevant) change of scale of $E$, is striking.

\subsection{Calibration of the model}

\begin{figure}[t]
    \centering
    \includegraphics[width=\columnwidth]{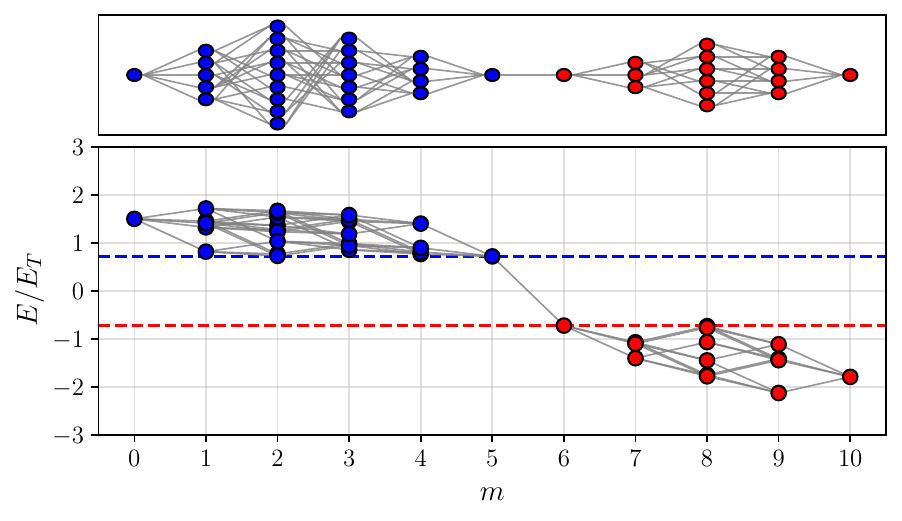}
    %\vskip-20pt
    \caption{Same representation as in Fig.~\ref{fig:1}d, using an instance of the calibrated model with Gaussian $P(h)$ instead of the experimental data. The lower panel reports the value of $E/E_T$ for each functional variant as a function of the distance from the blue reference variant (here with $E^{\rm {ref}}_B/E_T\approx 1.50$ and $E^{\rm {ref}}_R/E_T\approx-1.79$). The threshold value for which at least a mutational path remains is $E_C/E_T\approx0.72$ (dashed lines) and when normalized with respect to the largest of the two reference genotypes it reads $E_C/E^{\rm {ref}}_{\max}\approx0.40$.}
    \label{fig:2}
\end{figure}

\begin{figure*}[t]
    \centering
    \includegraphics[width=0.8\textwidth]{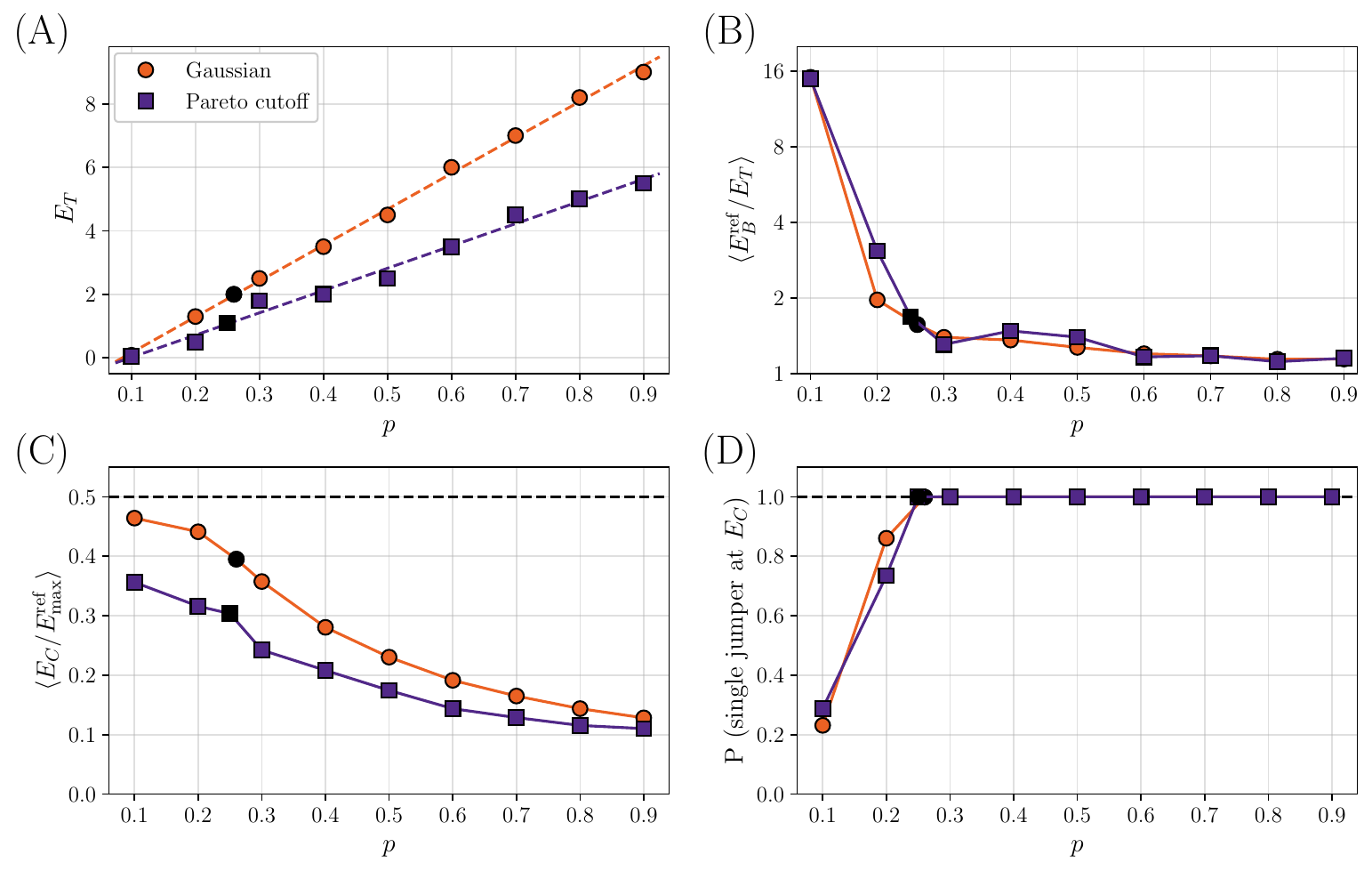}
    %\vskip-20pt
    \caption{Calibration of the two models with different $P(h)$, a Gaussian distribution or a Pareto distribution with cutoff. The black symbols correspond to the choice of $p$ after calibration.
    (A)~Value of $E_T$ for which $\langle M \rangle \approx 8$ as a function of $p$. Dashed lines are linear fits. 
    (B)~Average value $\langle E^{\rm {ref}}_B/E_T \rangle$ as a function of $p$. 
    (C)~Average of $E_C/E^{\rm {ref}}_{\max}$ as a function of $p$.
    (D)~Probability of having a single jumper at $E_C$, or equivalently that $E_C < E^{\rm {ref}}_{\min}$, as a function of $p$.
    }
    \label{fig:3}
\end{figure*}

The first step is to calibrate the parameters $(L,p,E_T)$ in such a way that our stylized model reproduces the basic phenomenology observed in Refs.~\cite{poelwijk2019learning,alqatari2024epistatic}. 
The requirements are the following.
\begin{itemize}
    \item The phenotype of the two reference sequences should be comparable to $E_T$, i.e. $|E^{\rm {ref}}|/E_T \simeq  1$, in order for $E_T$ to be a meaningful fitness scale.
    \item We want to reproduce a `bottlenecked' structure of the space of evolutionary paths connecting the two reference variants, as shown in Figs.~\ref{fig:1}d,~\ref{fig:2} and in Ref.~\cite{alqatari2024epistatic}.
    This is characterized by a single `jumper' genotype through which all paths must pass. 
    The mutation responsible for the  functionality switch occurs just after (or just before) this genotype. 
    We note that the jumper genotype can be connected by a single mutation to only one, or more than one, genotypes carrying the other phenotype.
    \item The `jumper' genotype should occur approximately at half distance along the evolutionary trajectory, as it is observed in experimental data (Fig.~\ref{fig:1}d).
    Furthermore, we want the largest possible number of functional intermediate variants to survive before and after the jump, thus maximizing the number of functional evolutionary paths connecting the two reference variants.
    \item The fitness variation produced by the `jumper' mutation(s) should be as large as possible, in such a way that $E_C$ is as close as possible to $E^{\rm {ref}}$. 
    If this is the case, a path can go in a single jump from having fitness close to the red reference, to having fitness close to the blue reference. Otherwise, one would necessarily have to tolerate a decrease in fitness to perform the functional switch.
    This makes the resulting mutational paths more viable for evolution.   
    \item Finally, for practical reasons, the number of mutations $M$ connecting the two reference variants should be small enough, otherwise enumerating all the $M!$ paths becomes computationally very expensive. 
    To avoid this, we impose $\langle M \rangle \sim 8$, a value close to that chosen in Ref.~\cite{alqatari2024epistatic}.
\end{itemize}

An important remark is in order about the expected value of $E_C$. Consider for illustration the case in which the tuning procedure reaches the target with a single greedy step for both colors. We can thus expect $E^{\rm {ref}}_B \sim \max(h_i)$ and $E^{\rm {ref}}_R \sim \min(h_i)$. In absence of network epistasis, the largest jumper mutation that can happen along a path has $|\Delta E| \sim \max(|h_i|) \sim E^{\rm {ref}}_{\rm max}$, but this mutation must bring the system from $E_C$ to $-E_C$, hence $2E_C \sim |\Delta E| \sim E^{\rm {ref}}_{\rm max}$. Based on this simple argument, the largest possible value of $E_C$ we can expect is such that $E_C/E^{\rm {ref}}_{\rm max} \sim 1/2$, as we observe numerically (Fig.~\ref{fig:3}c).

To fix the model parameters $(L,p,E_T)$ we proceed as follows.
First of all, we fix $L=500$ as the length of a typical protein. 
We checked that the results are very weakly dependent on $L$ (see the SI).
Next, for each $p$, we progressively increase $E_T$ until the average number of mutations $\langle M \rangle$ separating the two reference variants is $\langle M \rangle \sim 8$.
The resulting $E_T(p)$ is plotted as a function of $p$ for both choices of $P(h)$ in Fig.~\ref{fig:3}a, and we find that it is approximately linear in $p$. 
The linear fit allows us to fix $E_T(p)$ for all $p$.

The next step is to calibrate $p$. For this, we consider the ratio $|E^{\rm {ref}}|/E_T$ for both colors, which we want to be close to one. 
The evolution of its statistical mean with $p$ is shown in Fig.~\ref{fig:3}b.
For small $p$, we mostly perform random steps in the training, and the requirement that $M$ is small forces $E_T$ to be small as well (Fig.~\ref{fig:3}a). 
As a result, it is very likely to `overshoot' the target and end up with $|E^{\rm {ref}}|/E_T \gg 1$.
Hence, in order to have $|E^{\rm {ref}}|/E_T \sim 1$, we need a large enough $p$.

Next, we consider the connectivity threshold $E_C$.
We recall that $E_C$ is obtained, for each instance, in such a way that at least one single-mutational path connecting the two reference variants exists, with all intermediate states having $|E|>E_C$.
Because both reference sequences must always be included in the space of allowed genotypes,
increasing the value of $E_C$ beyond $E^{\rm {ref}}_{\min} = \min(E^{\rm {ref}}_B,|E^{\rm {ref}}_R|)$ is not meaningful~\cite{alqatari2024epistatic}.
This gives an upper bound for the largest value of $E_C$ for each instance.
We show in Fig.~\ref{fig:3}c its average value as a function of $p$, where the average is restricted to instances such that $E_C < 0.9 E^{\rm {ref}}_{\min}$.
When $p$ is too large, the tuning procedure requires too many greedy steps and as a consequence $\langle E_C/E^{\rm {ref}}_{\rm max} \rangle$ tends to decrease, contrarily to what we want in order to have bottleneck topologies with large~$E_C$. This suggests to lower the value of $p$ as much as possible.

However, in Fig.~\ref{fig:3}d we report the probability of having a single jumper at $E_C$, which we estimate for numerical convenience by the probability that $E_C < 0.9 E^{\rm {ref}}_{\min}$, as a function of $p$. When $p$ is too small, this probability drops quickly below one. 
This is because $|E^{\rm {ref}}|/E_T$ starts increasing, the reference variants are
obtained by many random steps, and the bottleneck structure is lost.

Taken together, these results suggest to choose the lowest value of $p$ such that the probability of a single jumper is still very close to one, which also gives both $E^{\rm {ref}}/E_T\sim 1$ and $E_C/E^{\rm {ref}}_{\rm max} \sim 0.5$. The value of $E_T(p)$ is fixed by the requirement that $\langle M \rangle \sim 8$. For both choices of $P(h)$, this optimal value is around $p=0.25$, although slightly larger values could be considered as well.

The crucial insight from the calibration is that one needs to tune $p$ in order to achieve a good balance between greedy and random steps. 
Correspondingly, as we will discuss later, the distribution $P(\tilde h)$ of the SMEs that are selected by the tuning procedure that generates the reference sequences (that may be interpreted as those `fixed by evolution') is very heterogeneous, featuring a coexistence of large SMEs (generated by greedy steps) and small SMEs (generated by random steps).

\subsection{Results for the calibrated model}

\begin{figure*}[t]
    \centering
    \includegraphics[width=\textwidth]{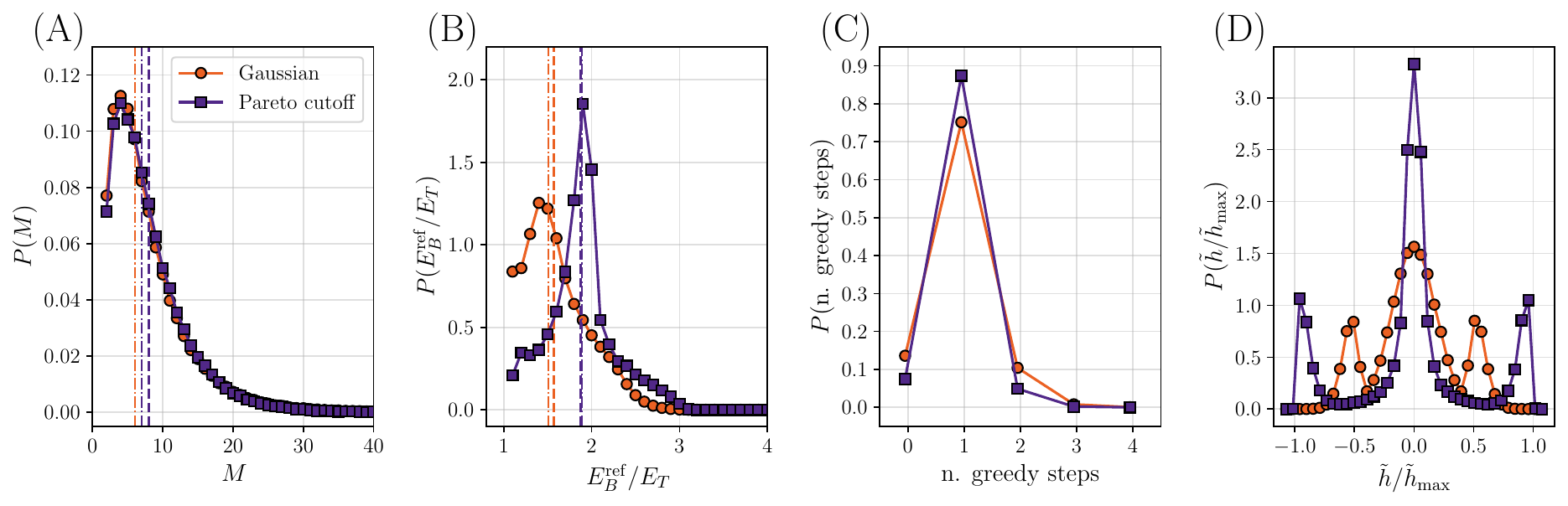}
    \caption{Distribution of some relevant quantities for two different calibrated models, the Gaussian model and the Pareto cutoff model. 
    In the former, $L=500$ SMEs are extracted from a Gaussian distribution with unit variance and the tuning procedure has $p=0.26$, $E_T=2.0$; in the latter, $L=500$ SMEs are extracted from a distribution with Pareto tails decaying with $\alpha=0.7$ and a cutoff $|h|<2.0$, and the tuning procedure has $p=0.25$, $E_T=1.1$.
    In each plot the dashed and dot-dashed vertical lines represent the mean and the median of the data, respectively. 
    (A)~Distribution of the number of mutations. 
    (B)~Distribution of the value of the positive (blue) reference phenotype divided by $E_T$. 
    (C)~Distribution of the number of greedy steps needed to reach the target value $E_T$ when generating the two reference variants.
    (D)~Distribution of SMEs $\tilde{h}_i$ selected by the tuning procedure to generate the two reference variants.}
    \label{fig:4}
\end{figure*}

In the following, we choose $L=500$ and the parameters corresponding to the black dots in
Fig.~\ref{fig:3} as the final calibrated parameters for which we report more detailed results: Gaussian $P(h)$ with $p=0.26$ and $E_T =2.0$ for the tuning procedure, and Pareto cutoff $P(h)$ with $p=0.25$ and $E_T=1.1$.  
Keeping these two models fixed, we generated many ($\sim 20000$) random instances of the model and studied the distribution of the relevant quantities that characterize the topology of the fitness landscape. 
An example of a `good' topology, closely resembling the one found experimentally, has already been given in Fig.~\ref{fig:2}. We want to quantify the probability of generating such a topology in the stylized model.

\begin{figure*}[t]
    \centering
    \includegraphics[width=\textwidth]{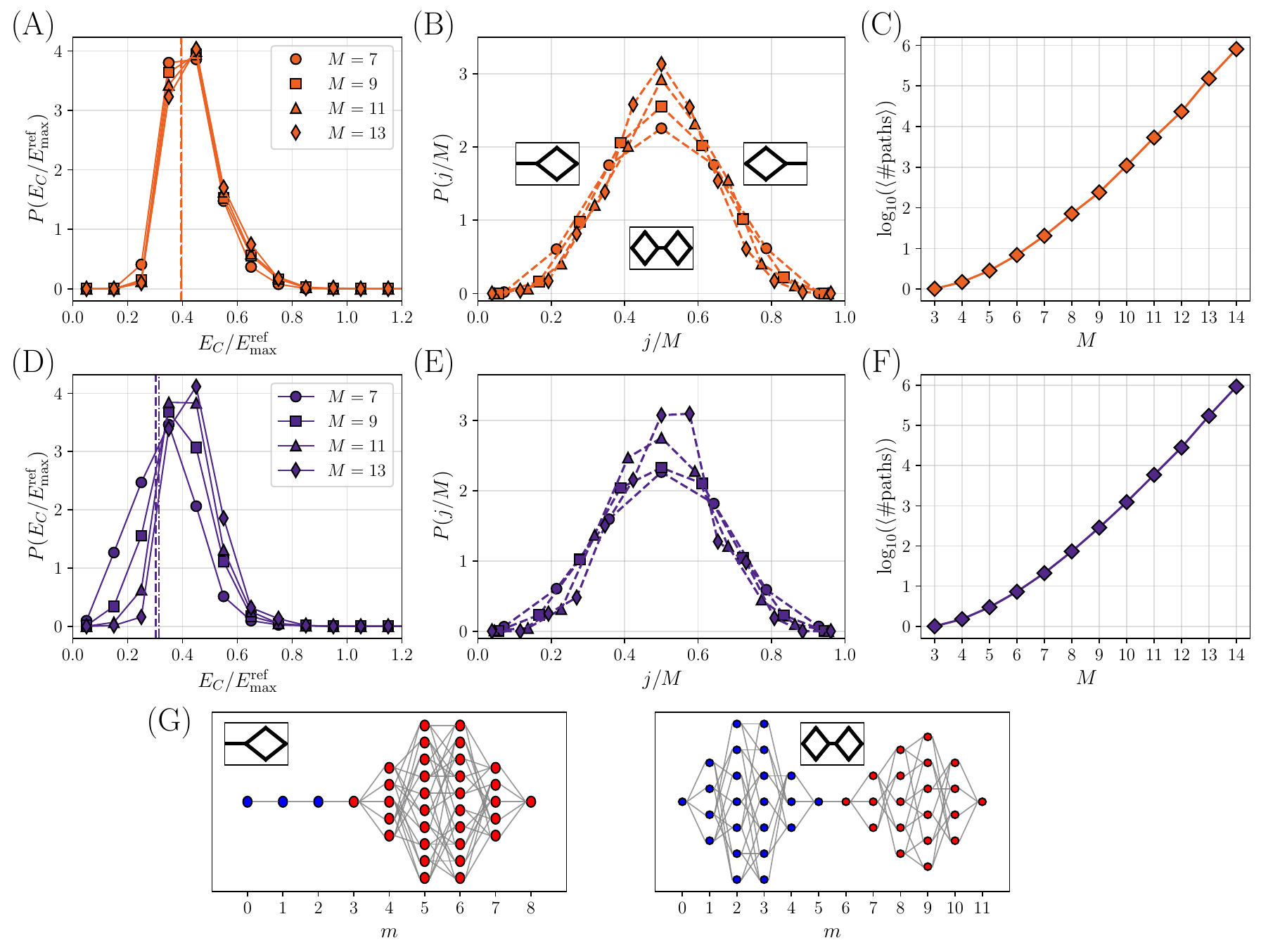}
    \caption{
    Statistical properties of the space of paths for two different calibrated model, the Gaussian model (A-C) and the Pareto cutoff model (D-F), with the same parameters as in Fig.~\ref{fig:4}.
    (A),~(D)~Distribution of values of $E_C/E^{\rm {ref}}_{\rm max}$, separately for several values of $M$. The mean and the median (dashed and dot-dashed vertical lines, respectively) are computed with respect to the full distribution (with all $M$ values).  
    (B),~(E)~Probability distribution of the jumper position $j$ along the path, for different values of $M$. 
    (C),~(F)~Average of logarithm of the number of paths that remain viable at $E_C$, as a function of $M$.
    (G)~Some examples of topologies, each associated to a schematic representation used in panel b.
    }
    \label{fig:5}
\end{figure*}

We first focus on the properties of the two reference variants, shown in Fig.~\ref{fig:4} for both models. 
More specifically, Fig.~\ref{fig:4}a shows the histograms of $M$, the total number of mutations separating the two reference sequences. 
We note that, by construction, $M \geq 2$, and we fixed $\langle M \rangle =8$.
We see that the distribution is quite broad, leading to a significant fraction of instances with $M$ as large as 20, and resembles that shown for elastic networks in Ref.~\cite{alqatari2024epistatic}.
However, when we investigate the mutational paths, we restrict the analysis to instances with $M\leq 14$ for computational tractability, as in Ref.~\cite{alqatari2024epistatic}. 
Fig.~\ref{fig:4}b shows the histograms of $E^{\rm {ref}}_B/E_T$ for the blue reference variant (similar results are obtained for the red reference variant due to the inherent symmetry of the model). 
We recall that, by construction of the training procedure, $E^{\rm {ref}}_B/E_T \geq 1$.
We observe that the histogram is peaked slightly above the minimal value,
which confirms that the tuning procedure produces $E^{\rm {ref}}_B\sim E_T$ and $E^{\rm {ref}}_R \sim -E_T$.
Fig.~\ref{fig:4}c shows the histograms of the number of greedy steps in the tuning procedure, which is peaked around one, as expected. Finally, Fig.~\ref{fig:4}d shows the histograms of the $M$ SMEs $\tilde h_i$ that are selected by the tuning procedure. These mutations characterize  {\it a posteriori} (after tuning) the space of intermediates between the reference variants, and can be thought of as being those `fixed' by evolution. We observe a strongly bimodal distribution $P(\tilde h)$ for both input distributions $P(h)$, which reflects the choice of the tuning procedure: mutations selected by the random steps tend to be small, while mutations selected by the greedy steps tend to be much larger (in a way that can be quantified by extreme value statistics, see the SI).

Next, we analyze the statistical properties of the space of paths connecting the reference variants (Fig.~\ref{fig:5}).
We recall that in the calibrated models we find $E_C < E^{\rm {ref}}_{\rm min}$ with probability one.
The space of paths thus contains a single genotype through which all paths must pass: raising $E_C$ would remove this genotype from the set, and disconnect the reference variants. 
A bottleneck is thus observed. 
We note that the jumper genotype, in most cases, is connected by a single mutation to only one genotype carrying the other phenotype, as in Fig.~\ref{fig:2}.
However, in a few cases, we also observe situations where the jumper genotype is connected to a few genotypes carrying the other phenotype.
In Fig.~\ref{fig:5}a,d we show the histograms of $E_C/E^{\rm {ref}}_{\rm max}$, separately for instances with given $M$. 
We observe that the histogram is almost independent of $M$ (with more variability observed for the Pareto cutoff model), and peaked around $E_C/E^{\rm {ref}}_{\rm max} \sim 1/2$ as discussed above.
Fig.~\ref{fig:5}b,e shows the histogram of the location of the jumper, called $j$, along the mutational path.
We observe that the jumper is often located mid-way between the reference variants, leading to a peaked histogram around $j/M \sim 1/2$. The peak (slowly) becomes sharper upon increasing $M$.
The qualitative shape of the possible topologies produced by the model is indicated in Fig.~\ref{fig:5}g.
Finally, we can compute the number of paths that remain viable at $E_C$.
The average of the logarithm of this number is shown in Fig.~\ref{fig:5}c,f as a function of $M$. 
We find that it grows almost linearly at large $M$, indicating that the number of paths grows exponentially in $M$. 
This suggests that there are exponentially many ways to reach the bottleneck, which could thus maintain a good evolutionary accessibility from both reference sequences~\cite{alqatari2024epistatic}.

We conclude that our stylized model, which features a global non-linearity on top of an underlying additive phenotype, with properly calibrated parameters, generates topologies that are very close to those observed experimentally in Ref.~\cite{poelwijk2019learning} and Sec.~\ref{sec:exp}.
An example is shown in Fig.~\ref{fig:2}, to be compared with the result obtained re-analyzing the experimental data from Ref.~\cite{poelwijk2019learning}, shown in Fig.~\ref{fig:1}d.
The model generates an ensemble of random fitness functions~\cite{Pahujani2025}, whose topology is likely (with probability one) characterized by a bottleneck, i.e., by a genotype through which all paths connecting the two reference variants must pass. 
Of course, the bottleneck can be made less pronounced if one allows for less stringent selection, i.e., reduces $E_C$~\cite{alqatari2024epistatic}. 
The bottleneck is more likely found mid-way along the path. 
These features are essentially independent of $M$, i.e., the number of mutations separating the two reference variants, at least in the limited range investigated here.

We note that there is quite a lot of flexibility in the choice of parameters.
The initial size $L$ can be varied in a broad interval without affecting much the results (see SI).
The tuning parameter $E_T$ has been chosen here in such a way to obtain the desired average of $M$, but other choices are possible.
The value of $p$, i.e., the probability of making a greedy step when constructing the reference variants, can be varied in a relatively wide range, provided it is neither too small (otherwise $E_T$ is too small and the reference phenotypes end up overshooting the fitness threshold) nor too large (otherwise $\langle E_C/E^{\rm {ref}}_{\rm max}\rangle$ decreases).
Finally, the {\it a priori} distribution $P(h)$ that characterizes the input distribution of SMEs can be varied essentially at will in a broad range (see also the SI for other choices of the input distribution). 
The tuning procedure used to construct the reference variants ensures that the SMEs selected {\it a posteriori} are broadly distributed, featuring the right balance between almost neutral SMEs and strongly beneficial or deleterious SMEs (Fig.~\ref{fig:4}d), which is a necessary condition for the emergence of a bottleneck.
We thus believe that our results are robust and do not depend crucially on the specific choices made during calibration.

\section{Discussion}

In this work, we have demonstrated that functional bottlenecks in protein fitness landscapes can emerge even in the absence of network epistasis, solely due to a non-linear mapping between an underlying additive trait and fitness.
To do so, we constructed and analyzed a simple stylized model of global epistasis, in which two reference variants are constructed from a common ancestor by selecting at random either a very beneficial mutation, or a neutral one, with the goal of acquiring two distinct new phenotypes.
We have shown that the emergence of bottlenecks separating the two phenotypes is closely tied to the heterogeneity of the single-mutation effects selected during the construction of the reference variants.
When mutational effects are heterogeneously enough, evolutionary paths between distinct phenotypes are constrained, leading to the formation of narrow evolutionary corridors.

Our findings provide a novel perspective on how fitness landscape topologies arise and suggest that strong constraints on evolutionary accessibility can exist even in systems governed by simple, global epistatic interactions~\cite{de2014empirical,Tenaillon2014,Pahujani2025,Hwang2017,otwinowski2018biophysical,otwinowski2018inferring,poelwijk2019learning,reddy2021global,phillips2021binding,johnson2023epistasis,buda2023pervasive}. 
In fact, in real evolution experiments, one always has access to a single realization of the process. Our stylized model, following and simplifying the approach of Ref.~\cite{alqatari2024epistatic}, constructs an ensemble of fitness landscapes and allows us to assess to what extent the realizations observed in a given experiment are `typical’ of a reasonable ensemble of possibilities.
Moreover, despite being stylized, our model successfully captures key qualitative features observed in experimental data, supporting the idea that broad mutational effect distributions play a central role in shaping evolutionary trajectories. 

Our results indicate that functional bottlenecks do not necessarily entail network epistasis. Such bottlenecks may instead arise because proteins, under strong evolutionary pressure to acquire new functions, tend to accumulate a few mutations that confer large benefits for the new function, alongside a handful of random neutral mutations introduced by the inherent stochasticity of evolution.
Hence, functional bottlenecks
may not necessarily indicate complex, higher-order interactions but could instead be a natural consequence of non-linear selection on an underlying additive trait.

Of course, network epistasis does exist in proteins, as it has been shown by a variety of approaches~\cite{lunzer2010pervasive,olson2014comprehensive,figliuzzi2016coevolutionary,rivoire2016evolution,poelwijk2016context,cocco2018inverse,starr2018pervasive,russ2020evolution,ballal2020sparse,lyons2020idiosyncratic,miton2021epistasis,bakerlee2022idiosyncratic,vigue2022deciphering,chen2024understanding}.
Our stylized model might serve as a null model to assess the relevance of network epistasis effects in the analysis of future experimental data.

We also stress that in different kinds of evolutionary dynamics, for instance neutral space evolution under weak selection~\cite{fantini2020protein,stiffler2020protein,bisardi2022modeling}, mutations are selected in a different way and our stylized model would likely not apply. The existence of functional bottlenecks during this kind of evolutionary dynamics is less established and might depend on subtle correlations between large groups of interacting
mutations~\cite{rivoire2016evolution,DiBari2024,rossi2024fluctuations}.
A recent study of a combinatorial landscape separating naturally occurring bacterial transcription factor binding sites did not find a bottleneck~\cite{westmann2024entangled}.

Our model is simple enough that it could possibly be solved analytically, using techniques from probability and statistical physics~\cite{de2014empirical,Hwang2017,Pahujani2025}. For instance, the tuning procedure consists in a random walk in the space of mutations, and preliminary results suggest that the probability of the resulting reference traits $E^{\rm {ref}}_B$ and $E^{\rm {ref}}_R$ can be written in simple form. Making some analytical progress would
eliminate the limitation on the number of mutations $M$ separating the two reference variants.
Future work could explore this possibility to extend our approach to larger mutational spaces, incorporate additional biological constraints, and investigate the robustness of these results across different fitness functions and evolutionary pressures, possibly with more than two phenotypes.

It should be noted that numerous techniques for extracting genotype-trait-fitness relationships directly from data -- including modern machine learning frameworks -- have emerged in what is now a rapidly expanding field (see e.g.,~\cite{figliuzzi2018pairwise,wang2018computational,tubiana2019learning,russ2020evolution,trinquier2021efficient,rives2021biological,malbranke2023machine,wagner2024genotype,razo2025learning}). The present work adopts a fundamentally different approach. Rather than attempting to fit specific empirical datasets, we propose here a simple stylized model, designed to identify relevant qualitative features and universal constraints in a generic way.

By highlighting the role of mutational heterogeneity in the emergence of functional bottlenecks, our study contributes to a deeper understanding of the constraints shaping protein evolution and hopefully opens new directions for theoretical and experimental exploration of evolutionary landscapes.

\section{Methods}

For convenience, we summarize here in closed form the main steps that have to be followed to generate an instance of our model.

\subsection{Genotype representation and underlying additive phenotype}

The first step consists in choosing a value of $L$ and 
considering a genotype $\aaa = (a_1, a_2, \dots, a_L)$ as a binary sequence of length $L$, where each site $i$ can take values $a_i \in \{0,1\}$. The genotype-dependent underlying additive trait is given by:
\begin{equation}
    E(\aaa) = \sum_{i=1}^{L} h_i a_i,
\end{equation}
where the single-mutation effects (SMEs) $h_i$ are independent and identically distributed according to the {\it a priori} distribution $P(h)$.
The two choices we used are a Gaussian distribution with zero mean and unit variance, and a Pareto distribution with cutoff:
\begin{equation}
    P(h) = A \begin{cases} 
        1\ , & |h| < 0.1 \ , \\
        1/|h|^{\alpha+1}\ , & 0.1 < |h| <2  \ ,\\
        0 \ , & |h|>2 \ ,
    \end{cases}
\end{equation}
with $\a=0.7$ and the constant $A$ determined by normalization.
After sampling, the SMEs are shifted to ensure $\sum_{i=1}^{L} h_i = 0$.
These choices fix an additive constant and an overall multiplicative factor in $E(\aaa)$, which can be absorbed by $\beta$ and $E_T$ in the fitness function, without loss of generality:

\subsection{Fitness function and phenotypic classes}
The global epistatic fitness function is a non-linear transformation of $E(\aaa)$,
defined separately for the blue (B) and red (R) phenotypes:
\begin{equation}
    F_{B}(E) = \frac{\phi_0}{1 + e^{\beta(E_\fth - E)}}, \quad F_{R}(E) = \frac{\phi_0}{1 + e^{\beta(E_\fth + E)}},
\end{equation}
where $\phi_0=1$ without loss of generality, $E_\fth$ is the functionality threshold parameter, and $\beta \gg 1$ controls the sharpness of the fitness transition. Accordingly, genotypes with $E(\aaa) > E_\fth$ are functional for the ``blue'' phenotype, those with $E(\aaa) < -E_\fth$ are functional for the ``red'' phenotype, and those with $|E(\aaa)| \ll E_\fth$ are non-functional.

\subsection{Reference variants construction}
The red and blue reference variants, $\aaa_R$ and $\aaa_B$, are independently generated starting from the ancestral genotype $\aaa = (0,0,\dots,0)$ by sequentially introducing mutations. At each step, with probability $p$, the mutation with the largest contribution to $E$ in the desired direction is chosen (greedy step), and with probability $1-p$, a random site is mutated (random step). The process is iterated until $E^{\rm {ref}}_R = E(\aaa_R) < -E_T$ and $E^{\rm {ref}}_B = E(\aaa_B) > E_T$.

\subsection{Mutational paths and bottleneck characterization}
The mutational distance $M$ between $\aaa_R$ and $\aaa_B$ is defined as the number of differing sites. All $M!$ possible evolutionary paths between the reference variants are considered, and a value $E_C$ is determined such that at least one path of single mutations exists, maintaining $|E(\aaa)| > E_C$ at all intermediate steps. The distribution of $E_C$ and the likelihood of a single critical `jumper' mutation define the presence and severity of the bottleneck.

\subsection{Data and Code availability}
All simulations and analyses were performed using scripts available at Zenodo (DOI: 10.5281/zenodo.18131102).
Data were downloaded from previously published sources \cite{poelwijk2019learning, buda2023pervasive}. Specifically, raw sequencing data were retrieved from the Sequence Read Archive (SRA) under BioProject accession PRJNA560590.

\acknowledgements{We warmly thank Sid Nagel and Martin Weigt for providing very useful feedback and inspiration all along the development of this work, and Emily Hinds for very useful guidance in reproducing the data analysis from Ref.~\cite{poelwijk2019learning}.
This research has been supported by first FIS (Italian Science Fund) 2021 funding scheme (FIS783 - SMaC - Statistical Mechanics and Complexity) from MUR, Italian Ministry of University and Research and from the PRIN funding scheme (2022LMHTET - Complexity, disorder and fluctuations: spin glass physics and beyond) from MUR, Italian Ministry of University and Research.}

%\bibsplit[2]
%Use \bibsplit to split the references from the body of the text. Value "[2]" represents the number of reference in the left column (Note: Please avoid single column figures & tables on this page.)

% Bibliography
\bibliography{references_nourl,pnas-sample}

\clearpage
% \section*{Supplementary Material for \\ ``Functional bottlenecks can emerge from non-epistatic underlying traits''}
\onecolumngrid

\begin{center}
    \large\bfseries
    Supplemental Material for\\
    ``Functional bottlenecks can emerge from non-epistatic underlying traits''
\end{center}

\vspace{1cm}

\twocolumngrid

\setcounter{section}{0}
\setcounter{equation}{0}
\setcounter{figure}{0}
\renewcommand{\thesection}{S\arabic{section}}
\renewcommand{\theequation}{S\arabic{equation}}
\renewcommand{\thefigure}{S\arabic{figure}}

\section{Details on the analysis of experimental data}

\subsection{Data collection}
\label{par:S1}

The experimental measurements that were analyzed in the main text were obtained from the public repository associated with Ref.~\cite{poelwijk2019learning}.
These are raw sequencing data capturing the genotype variants present in the red and blue channel after two-color Fluorescence Activated Cell Sorting. 
Following the procedure detailed in Ref.~\cite{poelwijk2019learning} (see the original paper for details), sequences with low quality scores and noisy measurements were filtered out. 
The enrichment scores for each genotype \textbf{a} were computed counting the occurrence of each variant in the two channels normalized to its abundance in the input channel, i.e. before sorting. 
We further normalized the enrichment scores with respect to the ones of the reference variants, such that $F_R(\textbf{a}_R)= F_B(\textbf{a}_B)=1$, and afterwards removed global epistasis via a non-linear function 
\begin{equation}
    E_\ph = \phi^{-1}(F_\ph) = F_\ph^{0.44} \ ,
\label{eq:nonlin}
\end{equation}
as proposed by the authors of Ref.~\cite{poelwijk2019learning}, separately for the blue ($\ph=B$) and red ($\ph=R$) fluorescence.

\subsection{Results with non-normalized enrichment scores}
\label{par:S2}

\begin{figure*}[t]
    \centering
    \includegraphics[width=\textwidth]{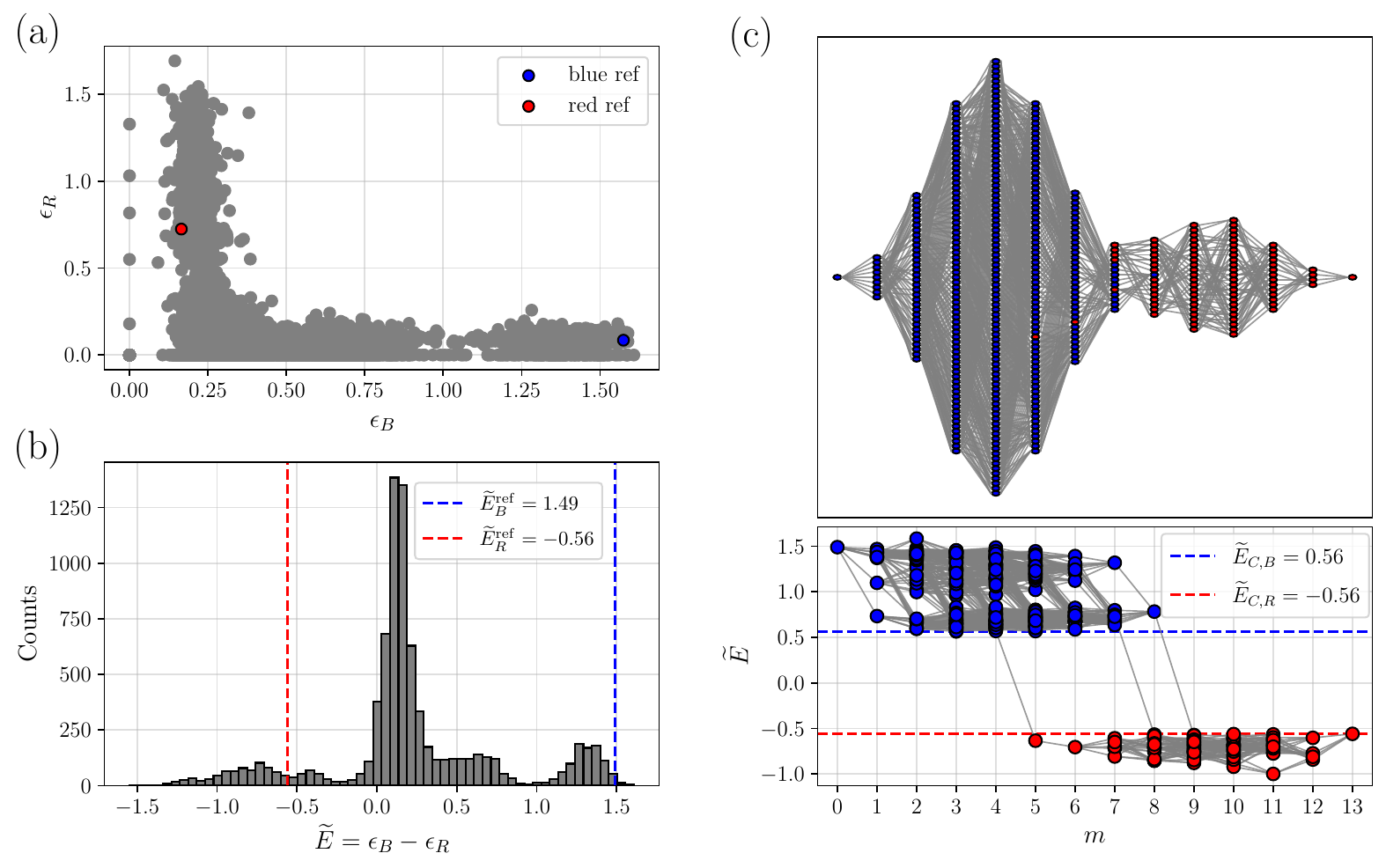}
    \caption{(a)~Scatter plot of the enrichment scores in the blue and the red channel, $\epsilon_B$ and $\epsilon_R$, not normalized with respect to the reference values. 
    For the blue reference variant $\epsilon_B = 1.58$,  $\epsilon_R = 0.09$. For the red one $\epsilon_B = 0.17$, $\epsilon_R = 0.73$. 
    (b)~Histogram of the values of $\widetilde{E}$ obtained for each of the $2^{13}$ variants. 
    The red and blue lines correspond respectively to the reference values $\widetilde{E}^{\rm {ref}}_R$ and $\widetilde{E}^{\rm {ref}}_B$.
    (c)~Topology of the space of paths obtained keeping only genotypes $\mathbf{a}$ with $|\widetilde{E}(\mathbf{a})|>\widetilde{E}_C$, where $\widetilde{E}_C=0.56$ (dashed lines) is the largest possible value such that the red and blue reference variants remain connected. The lower panel reports the values of $\widetilde{E}(\mathbf{a})$ for each functional genotype (blue dots for $\widetilde{E}>\widetilde{E}_C$ and red dots for $\widetilde{E}<-\widetilde{E}_C$) as a function of the number of mutations $m$ from the blue reference sequence, with the gray lines connecting pairs of genotypes that differ by a single mutation. The upper panel shows the resulting graph of connections.}
    \label{fig:SRama}
\end{figure*}

In the main text, and as discussed in Sec.~\ref{par:S1}, we normalized the enrichment scores to the reference values, a choice that was not made in the original Ref.~\cite{poelwijk2019learning}. 
Using normalized enrichment scores, we find that no path of single mutations exists that connects the two reference variants while always maintaining $|E(\textbf{a})|\geq E^{\rm {ref}}_{\max}$, hence we need to consider $E_{\rm{th}} < E^{\rm {ref}}_{\max}$ in order to preserve the connection between the reference variants, as discussed in the main text. 
Instead, Ref.~\cite{poelwijk2019learning} shows a topology of the space of paths connecting genotypes with $|E(\textbf{a})|\geq E_{\rm{th}}$ and $E_{\rm{th}} = E^{\rm {ref}}_{\rm{min}}$. This discrepancy arises entirely from the choice of not normalizing the phenotypic trait.

The space of paths presented in Ref.~\cite{poelwijk2019learning} is constructed by defining a phenotype $E = \sqrt{\epsilon^2_B + \epsilon^2_R}$, where $\epsilon_B$ and $\epsilon_R$ are the enrichment scores in the two channels, which are {\it not} normalized with respect to the enrichment scores of the red and blue reference variants. 
However, they are rescaled to reflect the experimental brightness ratio between the parental blue and red genotypes, which is $0.172$, and subsequently transformed using the non-linear function in Eq.~\ref{eq:nonlin} to remove global epistasis.
To carry out a proper comparison with our analysis, we reproduce the space of paths of Ref.~\cite{poelwijk2019learning} modifying their definition of phenotype to $\widetilde{E} = \epsilon_B - \epsilon_R$. 
Fig.~\ref{fig:SRama}a shows that this is a minor change: because $\epsilon_R\sim 0$ when $\epsilon_B > 0$ and vice versa, the two phenotypes 
are essentially exclusive and as a consequence
$E = \sqrt{\epsilon^2_B + \epsilon^2_R} 
\sim \max(\epsilon_B,\epsilon_R) \sim | \widetilde{E}|$.
Hence, $\widetilde{E}$ is numerically very close to the score used in the original Ref.~\cite{poelwijk2019learning}, while keeping the negative sign for the red phenotype. 

Compared to the normalized case shown in the main text, the resulting distribution of phenotypes, shown in Fig.~\ref{fig:SRama}b, is not symmetric around zero: the red reference sequence has $|\widetilde{E}^{\rm {ref}}_R| =  0.56$ and the blue one is more than twice as large, with $\widetilde{E}^{\rm {ref}}_B =1.49$. 
This asymmetry arises entirely from the absence of normalization.
When looking for the largest value of the functionality threshold $\widetilde{E}_{\rm{th}}$ such that at least one viable mutational path exists between the two reference variants, we find that $\widetilde{E}_C$ saturates to  $\widetilde{E}^{\rm {ref}}_{\rm{min}}$ and does not produce a topology with a single jump. 
The topology is the same as the one shown in Ref.~\cite{poelwijk2019learning}, and we consistently find that $\widetilde{E}_C$ = $\widetilde{E}^{\rm {ref}}_{\rm{min}}$ is the critical value above which any connection between the two reference sequences is lost.
This is exactly what we would expect given the imbalance between the two reference phenotypes.
However, given this imbalance, we believe that choosing a symmetric threshold, i.e. choosing the same $\widetilde{E}_{\rm{th}}$ for both functions, is not appropriate. 

We therefore believe that the choice of normalizing the enrichment scores to the reference variants provides a better view of the global fitness landscape, with the distribution of phenotypic values centered around zero. 
This choice results in a critical threshold $\widetilde{E}_C < \widetilde{E}_{\rm {ref}}$, as discussed in the main text.

\subsection{Removing network epistasis from the data}

\begin{figure*}[t]
    \centering
    \includegraphics[width=\textwidth]{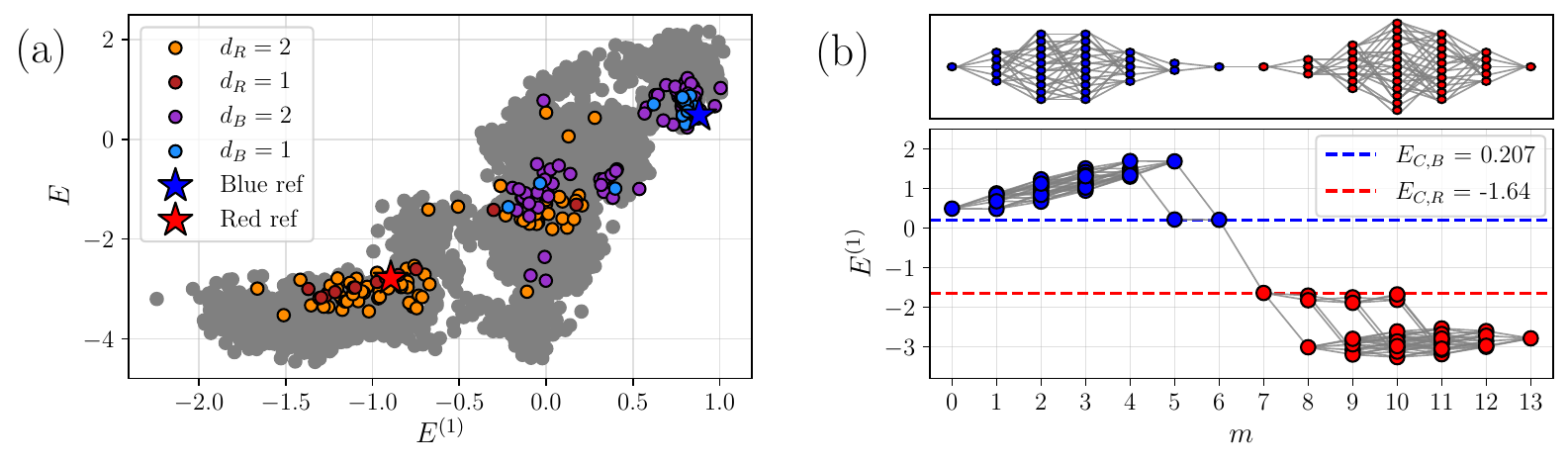}
    \caption{Linearization of the phenotypes according to the expansion described in Eq.~\eqref{eq:LinearizationAroundWT}. 
    (a)~Scatter plot of the phenotype values shown in the main text, $E$, and after the expansion, $E^{(1)}$. The Spearman correlation coefficient is about $0.7$.
    The red and blue stars correspond respectively to the red and blue reference variants. 
    (b)~Topology of paths surviving at the connectivity thresholds $E^B_C$ and $E^R_C$. 
    Here $E^{(1)}$ is the linearized additive phenotype.
    }
    \label{fig:S2}
\end{figure*}

From now on, we stick to the choice made in the main text, of normalizing the enrichment scores to the ones of the reference variants. 
We note that even after the non-linear fitness function has been inverted by the fitting procedure we described, the effect of epistasis is still present in the data.  
This is due to the fact that the fitness cannot be simply expressed as a nonlinear function of an additive trait. 
An interesting test is then to try and remove this second source of epistasis (network epistasis) and check whether one can still obtain a topology similar to the one discussed in Refs.~\cite{poelwijk2019learning,alqatari2024epistatic} and in the main text.
We do so by defining an additive phenotype $E^{(1)}_\ph$ obtained expanding the phenotypes $E_\ph$ around their respective reference variant $\textbf{a}_\ph$ and cutting the higher order terms, namely
\begin{equation}\label{eq:LinearizationAroundWT}
    \Enoep(\textbf{a}) = E_{\ph}(\textbf{a}_{\ph}) + \sum_{i=1}^{13} h^{\ph,(1)}_i (a_{\ph,i} - a_i) \ .
\end{equation}
After linearization, we combine the two phenotypes into a single additive phenotype $E^{(1)}(\textbf{a}) = E_B^{(1)}(\textbf{a}) - E^{(1)}_R(\textbf{a})$.

Fig.~\ref{fig:S2}a shows a scatter plot of the approximated additive phenotype $E^{(1)}(\textbf{a})$ against $E(\textbf{a})$.
In the figure we highlight the reference sequences (blue and red stars) along with the ones that are at a small Hamming distance from the blue ($d_B$) and red ($d_R$) variants.
One notices that $E$ and $E^{(1)}$ are not simply linearly related, even if they have a relatively high Spearman correlation coefficient of $0.7$.
This suggests that the first order expansion is not able to fully capture all mutational effects and that network epistasis is indeed present in the data~\cite{poelwijk2019learning}. 
We note that the variants at small Hamming distance from the reference ones display a better correlation than the very distant ones, as expected.

Keeping in mind that the linearization is only an approximation, we analyzed the space of viable paths connecting the two reference variants. 
Following the same approach described in the main text, we look for the largest value of $E_{\rm th}$ for which at least one path allowing for a functionality switch survives, and call this value $E_C$. 
However, because the two reference sequences in this case turn out to be quite asymmetric, with $E^{\rm {ref}}_B\sim 0.49$ and $E^{\rm {ref}}_R\sim 2.8$, we allowed for an asymmetric connectivity threshold~\cite{alqatari2024epistatic} 
for the paths joining the two variants (see also the discussion in Sec.~\ref{par:S2}).
Instead of increasing $E_{\rm th}$ symmetrically from 0 and impose $|E|>E_{\rm th}$ for every path, we impose $E>E^B_{\rm th}$ and $E<-E^R_{\rm th}$ separately for the two phenotypes.
The largest values for which a path survives are $E^B_C = 0.21$ and $E^R_C = 1.64$, and we obtain the topology shown in Fig.~\ref{fig:S2}b.
Note that in the end, $E^B_C/E^{\rm {ref}}_B\sim 0.42$ and $E^R_C/E^{\rm {ref}}_R\sim 0.59$, hence we always end up with a connectivity threshold at about half the reference phenotype.
We observe a significant reduction in functional variants, approximately at half distance between the two reference sequences, and the emergence of a bottleneck topology. 

\begin{figure}[b]
    \centering \includegraphics[width=\columnwidth]{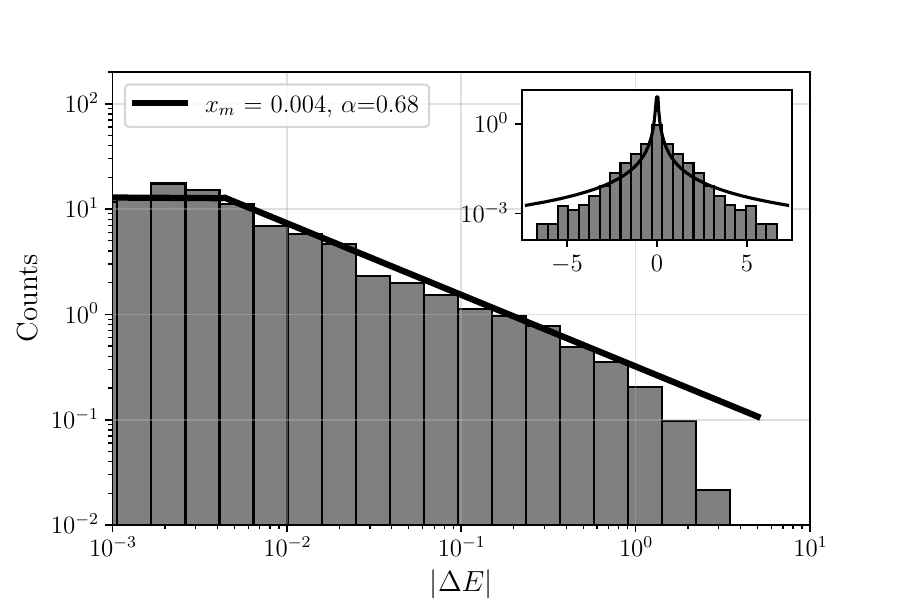}
    \caption{ Histogram of the absolute value $|\Delta E|$ for all SMEs extracted from the dataset of Ref.~\cite{buda2023pervasive}.}
    \label{fig:S3}
\end{figure}

\section{Analysis of additional data for single-mutation effects}

In Ref.~\cite{buda2023pervasive} an extensive review of combinatorial experiments similar to that of Ref.~\cite{poelwijk2019learning} is presented. 
The authors analyzed the results from ten studies (see references in~\cite{buda2023pervasive}) of combinatorially complete fitness landscapes of seven different enzymes, totaling 1440 genotype-phenotype data-points.
The experimental results for fitness are then translated into an underlying phenotype by attempting at removing non-linearity (see Ref.~\cite{buda2023pervasive} for details), and the authors provide the resulting distribution of SMEs that can be downloaded from the paper repository. It should be noted, however, that most of these landscapes connect a reference wild type to an evolved variant with increased fitness,
rather than two variants with distinct functionality.

In Fig.~\ref{fig:S3}, we show the same distribution as in Ref.~\cite{buda2023pervasive}, plotted similarly to the other figures in this paper.
Once again the fit with the Pareto distribution seems to perform well, although with different values of $x_m$ and $\alpha$. 
Notice that the uniform behavior at small $|\Delta E|$, as well as the cutoff at large $|\Delta E|$, are not as sharp as in the data from Ref.~\cite{poelwijk2019learning}. 
This is probably due to the fact that Ref.~\cite{buda2023pervasive} combined results for many different experiments, which have different sensitivity and a different range of measurable fitness.

\begin{figure*}[t]
    \centering  \includegraphics[width=\textwidth]{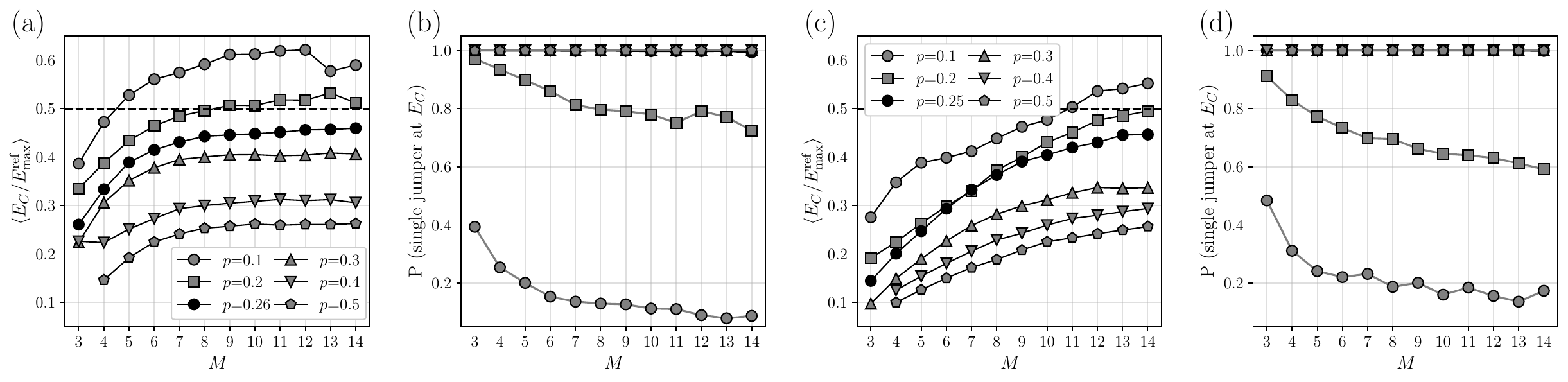}
    \caption{Statistical properties of the space of paths upon varying the number of mutations $M$, at fixed values of $p$, for the Gaussian (a),(b), and the Pareto cutoff (c),(d) distributions. 
    (a),(c)~The average value of $E_C/E^{\rm {ref}}_{\max}$.
    (b),(d)~The fraction of single jumpers present at $E_C$.
    }
    \label{fig:S5}
\end{figure*}

\begin{figure*}[t]
    \centering
    \includegraphics[width=\textwidth]{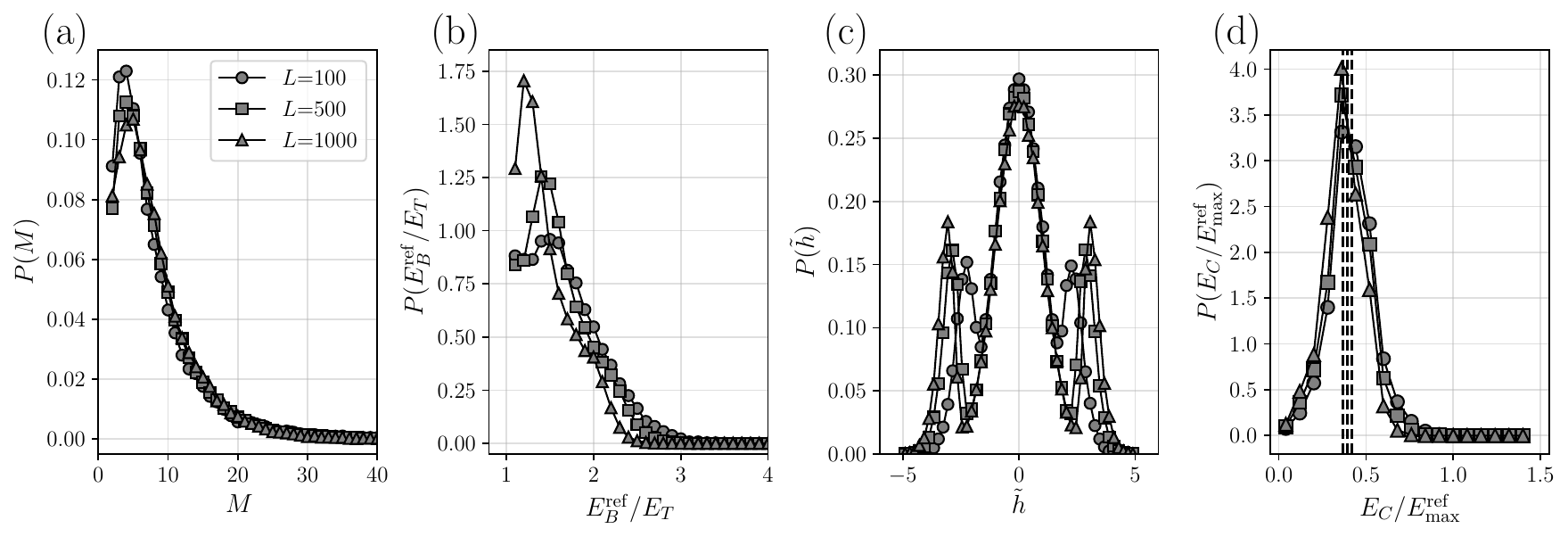}
    \caption{Statistics of a few relevant quantities for the model with Gaussian distribution, for $L=100,500,1000$. 
    (a)~Distribution of the number of mutations. 
    (b)~Distribution of the value of the positive (blue) reference phenotype divided by $E_T$.
    (c)~Distribution of the selected SMEs $\tilde{h}$.
    (d)~Distribution of $E_C/E^{\rm {ref}}_{\max}$.
    }
    \label{fig:S4}
\end{figure*}

\section{Additional results on the stylized model}

\subsection{Evolution of the space of paths with the number of mutations}

In the main text we presented the statistics of $E_C/E^{\rm {ref}}_{\max}$ for the calibrated models, i.e. for the parameter choices $p$ and $E_T$ that maximize both the connectivity threshold  $E_C/E^{\rm {ref}}_{\max}$ and the probability of observing bottleneck structures. 
These two quantities are shown in Fig.~\ref{fig:S5} as a function of the number of mutations $M$ separating the reference variants, for different values of $p$ and $E_T$. 
As explained in the main text, the dotted line at $E_C = 0.5 E^{\rm {ref}}_{\max}$ corresponds to the maximum value of connectivity threshold that can be found when observing a bottleneck with a single functionality switch. 
At large $p$, it can be seen that although the fraction of bottlenecks remains equal to one, $\langle E_C/ E^{\rm {ref}}_{\max} \rangle$ decreases with increasing $p$. 
Low values of $E_C/ E^{\rm {ref}}_{\max} $ correspond to a connectivity threshold that lies deep within the region of non-functionality, making the corresponding evolutionary trajectories unlikely to occur.
On the other hand, for small $p$, corresponding to a random selection of mutational effects, the fraction of observed bottlenecks decreases significantly. 

For the values of $p$ and $E_T$ chosen after calibration (black dots), we observe simular results to Ref.~\cite{alqatari2024epistatic}. The probability of a bottleneck remains always equal to one, while the value of $E_C/E^{\rm {ref}}_{\max}$ increases with $M$ and saturates to the expected maximum.

\begin{figure*}[t]
    \centering
    \includegraphics[width=\textwidth]{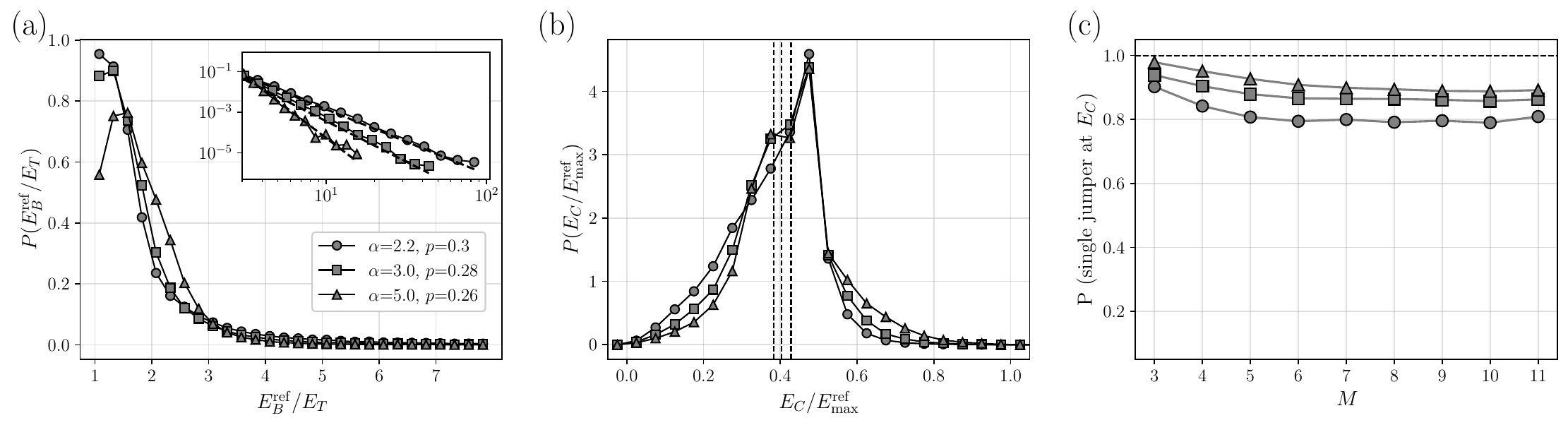}
    \caption{Distribution of the relevant quantities for the model with Pareto distributed $h$ values. Results are shown for different exponents $\alpha = 2.2$, $3.0$, and $5.0$.
    (a)~Distribution of $E^{\rm {ref}}_B/E_T$. Inset: tails of the distributions in log-log scale. The dashed lines are obtained fitting with an exponent equal to $\alpha$.
    (b)~Distribution of the value of $E_C/E^{\rm {ref}}_{\max}$.
    (c)~Fraction of single jumpers at $E_C$ as a function of the number of mutations $M$.
    }
    \label{fig:S6}
\end{figure*}

\subsection{Dependence on system size}

All the results reported in the main text have been obtained by fixing the number of SMEs to $L=500$. 
This would correspond to a protein of $500$ amino acids or to an elastic network with $500$ edges. 
To check the consistency of our results we also perform simulations varying the length $L$ of the binary sequence. 
In particular we study $L=100$ and $L=1000$ as well.
We focus on the same parameters used in the main text, namely a Gaussian with $\mu=0$, $\sigma=1.0$ and we determine $p$, $E_T$ fixing $\langle M \rangle \approx 8$. 
As expected, the required value of $E_T$ increases with $L$, because the maximum of the $h_i$ increases proportionally to $\sqrt{\log L}$ as dictated by extreme value statistics. The growth is slow enough, however, and it is compensated by the tuning of $E_T$, in such a way that
the results discussed in the main text are not significantly affected by the choice of the system size.
In Fig.~\ref{fig:S4}, panels (a) and (b), we plot the distribution of the number of mutations and of the value of $E^{\rm {ref}}_B/E_T$, respectively. 
We see that, with an appropriate choice of $E_T$ and $p$ for each $L$, the results are almost indistinguishable for the three values of $L$. 
In panels (c) we plot the distribution of the values of $h_i$ selected by the tuning procedure.
We see that, as $L$ increases, the peak of strongly beneficial (or strongly deleterious) mutations shifts towards larger absolute values, sharpening the separation between greedy and random steps.   
Finally, in panel (d) we show the distribution of $E_C/E^{\rm {ref}}_{\max}$, which is once again unaffected by the choice of $L$.

\begin{figure*}[t]
    \centering
    \includegraphics[width=\textwidth]{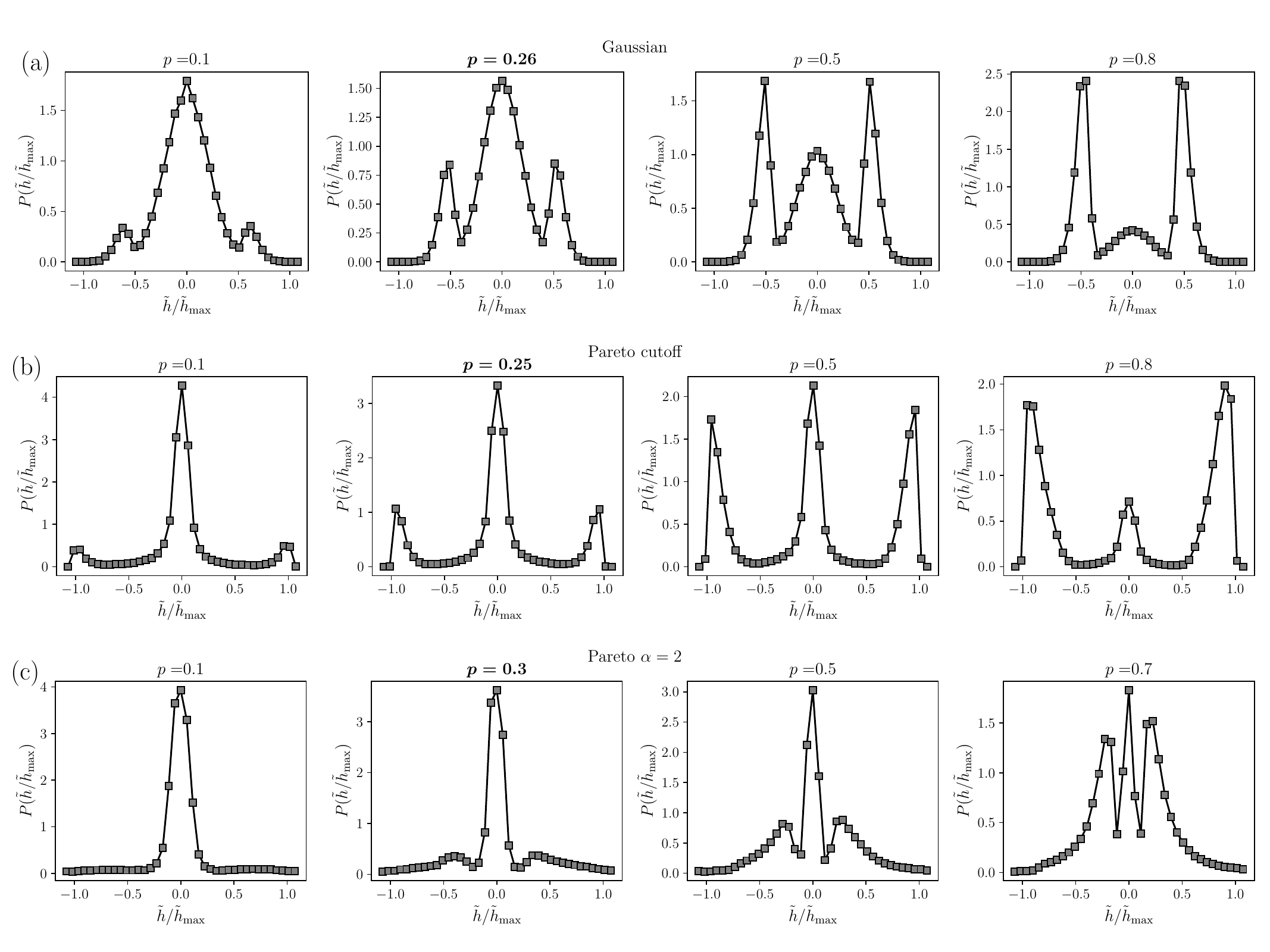}
    \caption{Distribution of $P(\tilde{h}/\tilde{h}_{\rm{max}})$. Comparison between calibrated (bold $\boldsymbol{p}$) and non calibrated models. 
    (a)~Gaussian input distribution $P(h)$. Calibrated model for $p=0.26$, $E_T = 2.0$.
    (b)~ Pareto cutoff distribution $P(h)$: $\alpha= 0.7$, $x_m=1$, $\epsilon = 2$. Calibrated model for $p=0.25$, $E_T=1.1$.
    (c)~ Pareto distribution $P(h)$ with $\alpha = 2.0$. Calibrated model for $p=0.3$, $E_T = 8.0$.
    }
    \label{fig:S7}
\end{figure*}

\subsection{Heavy tailed distribution of single-mutation effects}

The analysis of experimental data for SMEs suggests that the initial part of the distribution is described by a heavy-tailed Pareto distribution, which is then cutoff at large $|\Delta E|$, in part because of biological reasons (the effect of a mutation on fitness cannot be too large), in part due to the limited range of fitness variations that can be measured in experiments.

Yet, it is interesting to check what would happen in our model if the heavy tail would be allowed to continue up to infinity (keeping in mind that the finite value of $L$ will induce a cutoff anyways).
Hence, we show here some results obtained for various choices of the input distribution $P(h)$ with Pareto tails decaying according to an exponent $\alpha$ as in Eq.(5) of the main text, without cutoff.
We analyze different values of $\alpha=2.2, 3,5$, for which the variance of the distribution is finite.
We fix the system size to $L=500$ as in the main text, to facilitate comparison.
We then proceed as described in the main text and we characterize the statistics of $E^{\rm {ref}}_{\ph}/E_T$, $E_C/ E^{\rm {ref}}_{\max}$, and the topologies that arise from this input distribution of SMEs. 
The results are shown in Fig.~\ref{fig:S6}, for the optimal choice of $p$ and $E_T$.

For this choice of the parameters we find $\langle E_C/E^{\rm {ref}}_{\max}\rangle \simeq 0.5$, which is the largest possible value of $E_C$ maximizing the fraction of bottleneck topologies in the ensemble.
We note that if there is only one greedy step, the value of the reference phenotype $E_B^{\rm {ref}}$ is given by the sum of $M-1$ values drawn from $P(h)$ and a value drawn from $P^L_{\max}(h_{\max})$, the probability distribution of the maximum value of the $L$ variables $h$, which is easily derived from extreme value statistics. If $P(h)$ has heavy tails, we expect the sum to be dominated by the maximum and, as a consequence, $P(E^{\rm {ref}}_B)\sim P^L_{\max}(h_{\max})$.
The same reasoning works for the red reference phenotype $E^{\rm {ref}}_R$. 
In Fig.~\ref{fig:S6}a we show the plot of $E^{\rm {ref}}_B$ for different values of the exponent $\alpha$.
In the inset of this panel one can see how the tails of such distribution decay following a power law with the same exponent $\alpha$, as expected.

The distribution of $E_C/E^{\rm {ref}}_{\max}$, shown in Fig.~\ref{fig:S6}b, seems to be rather independent of the exponent $\alpha$, and is centered around $0.5$ as observed in the previous cases. 
What is different is the fraction of single jumpers at $E_C$, shown in Fig.~\ref{fig:S6}c. 
At variance with the Gaussian and Pareto cutoff cases, here the fraction of bottleneck topologies is different from one even for the optimal values of $p$ and $E_T$. 
This difference is due to the fat tails distribution of the reference phenotype $E^{\rm {ref}}_{\ph}$. 
Since the distribution is very broad, it is possible to have a very unbalanced situation in which one of the two reference phenotypes is much larger than the other (in particular, when $E^{\rm {ref}}_{\max} \gtrsim 2E^{\rm {ref}}_{\min}$), which allows for the saturation of $E_C$ to $E^{\rm {ref}}_{\min}$ and the absence of a bottleneck.
As expected, this effect disappears for larger values of $\alpha$, when the tails are less pronounced, and absent in the Gaussian and Pareto cutoff distributions.

\subsection{Balance between neutral and strongly non-neutral mutations}

A central finding of our work is that the emergence of functional bottlenecks is contingent upon a specific balance within the distribution of {\it a posteriori} `fixed' mutational effects, $P(\tilde{h})$. While the {\it a priori} distribution of available mutations, $P(h)$, can be relatively flexible, the realized mutational neighborhood of the reference genotypes must exhibit a high degree of heterogeneity to produce the desired topology. In Fig.~\ref{fig:S7}, we illustrate this balance by comparing calibrated models against non-calibrated ones. The calibration procedure essentially ensures that the two reference genotypes are separated by an underlying trait distance that allows for a `neutral bulk' of mutations punctuated by a sufficient frequency of large-effect mutations. As illustrated in Fig.~\ref{fig:S7}, when the model is not properly calibrated, the resulting distribution $P(\tilde{h})$ fails to achieve this critical composition in two distinct ways. 
If the fraction of large-effect mutations is too small, the fitness landscape remains overly smooth and accessible, preventing the formation of fitness valleys (at too small $p$).
Conversely, if too many large-effect mutations dominate the distribution (at too large $p$), the landscape becomes fragmented, and the selection pressure required to form a `corridor' between functional states must be too small.
These results substantiate our claim that functional bottlenecks do not emerge by default from any nonlinear mapping. Rather, they require a specific balance between a majority of nearly neutral mutations and a minority of strongly non-neutral ones, which is provided by our procedure to construct the reference genotypes.
This provides a clear example of why the construction is non-trivial: the bottleneck structure is a direct consequence of the heterogeneity in mutational effect sizes, and it vanishes when this heterogeneity is removed or incorrectly tuned. 
\begin{figure*}[t]
    \centering
    \includegraphics[width=\textwidth]{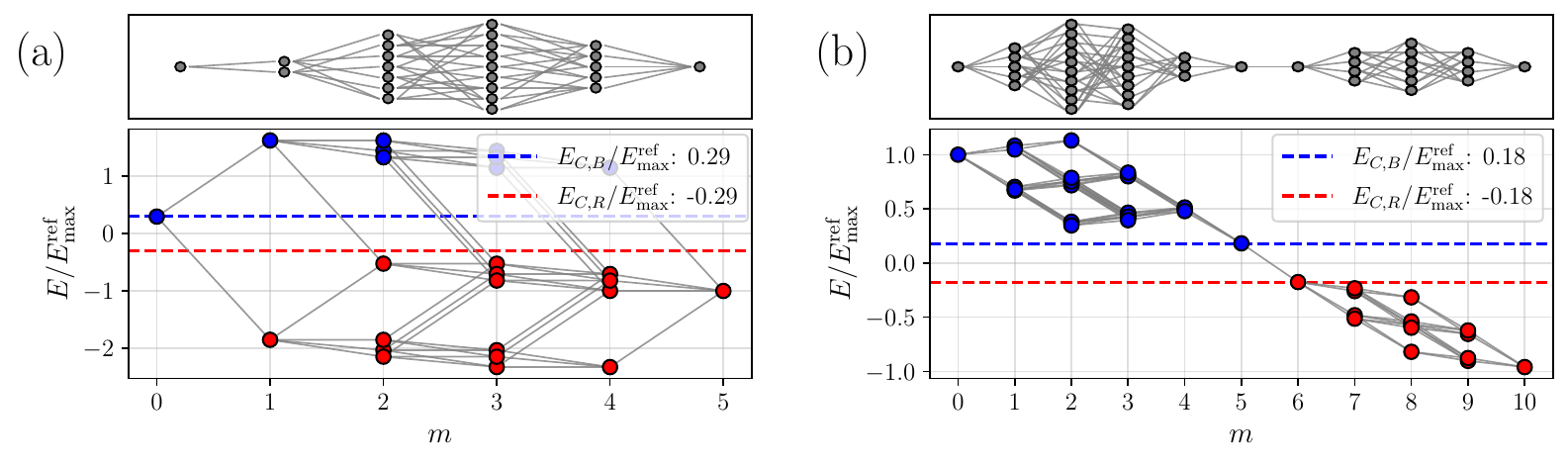}
    \caption{Topology of paths surviving at the connectivity threshold for two instances drawn from uncalibrated models. $E$ and $E_C$ are scaled with respect to the largest of the two reference genotypes $E^{{\rm{{ref}}}}_{\rm{max}}$.
    (a)~ Gaussian input distribution $P(h)$, $p=0.1$, $E_T = 0.08$. Note the typical behaviour: the two reference genotypes are unbalanced, $E^{{\rm{{ref}}}}_{B}/E_T = 2.73$, $E^{{\rm{{ref}}}}_{R}/E_T = -9.25$, $E^{{\rm{{ref}}}}_{B}/E^{{\rm{{ref}}}}_{R} = 3.39$  ; the connectivity threshold reaches its upper bound, which is the smallest between the two reference genotypes $E^{{\rm{{ref}}}}_{\rm{min}}$.
    (b)~ Gaussian input distribution $P(h)$, $p=0.8$, $E_T = 8.2$. Here the reference genotypes are symmetric and close to $E_T$:
    $E^{{\rm{{ref}}}}_{B}/E_T = 1.11$, $E^{{\rm{{ref}}}}_{R}/E_T= -1.06$, $E^{{\rm{{ref}}}}_{B}/E^{{\rm{{ref}}}}_{R} = 1.04$; there is a single jumper mutation at half distance but the connectivity threshold is very small.
    }
    \label{fig:S8}
\end{figure*}

In Fig.~\ref{fig:S8} we show typical examples of paths that emerge when the model is not properly calibrated.
As already stated before, if the fraction of large-effect mutations is too small, no single jumper mutation stands out and the functionality switch can occur at any step along the evolutionary path. This is illustrated in Fig.~\ref{fig:S8}a: the fitness landscape is smooth and no bottleneck emerges during evolution. Conversely, if large-effect mutations dominate the distribution, as in Fig.~\ref{fig:S8}b, the jump required to switch functionality is small: evolution has to accommodate substantial functionality loss with respect to the reference genotypes before being able to switch functionality, thus making these paths very unlikely to occur.

%\bibliography{pnas-sample,references_nourl}

\end{document}